\documentclass[aps,pre,twocolumn,longbibliography,showpacs,amsmath,amsmath,amssymb,amsfonts,notitlepage,superscriptaddress]{revtex4-1}

\usepackage{graphicx}
\usepackage{dcolumn}
\usepackage{bm}
\usepackage{hyperref}
\usepackage{xcolor}
\usepackage{graphicx}
\usepackage{dcolumn}
\usepackage{bm}
\usepackage{color}
\usepackage{multirow}
\usepackage[normalem]{ulem}
\usepackage{color}
\usepackage{physics}
\usepackage{bbm}
\usepackage{tikz-cd}
\usepackage{tikz}
\usepackage{algorithm}
\usepackage{algpseudocode}

\newcommand{\vx}{\bm{x}}
\newcommand{\vy}{\bm{y}}
\newcommand{\vz}{\bm{z}}
\newcommand{\vf}{\bm{f}}
\newcommand{\vg}{\bm{\gamma}}
\newcommand{\vLambda} {\bm{\Lambda}}
\newcommand{\vK}{\bf{K}}
\newcommand{\vxt}{\tilde{\vx}}
\newcommand{\xt}{\tilde{x}}
\newcommand{\xtm}{\tilde{X}}
\newcommand{\la}{\langle}
\newcommand{\ra}{\rangle}
\newcommand{\be}{\begin{equation}}
\newcommand{\ee}{\end{equation}}
\newcommand{\bea}{\begin{eqnarray}}
\newcommand{\eea}{\end{eqnarray}}

\newcommand{\tleft}{\mathcal{T}_0}
\newcommand{\tright}{\mathcal{T}_1}
\newcommand{\tcal}{\mathcal{T}}

\newcommand{\rev}[1]{#1}

\definecolor{commentgray}{HTML}{8D9B90}

\begin{document}


\title{\textbf{Making rare events typical in $d$-dimensional chaotic maps} 
}%

\author{Yllari K. González-Koda}
\email[]{yllari@ugr.es}
\affiliation{Departamento de Física Teórica y del Cosmos, Universidad de Granada, Granada 18071, Spain}

\author{ Ricardo Guti\'errez}
\email[]{rigutier@math.uc3m.es}
\affiliation{Universidad Carlos III de Madrid, Departamento de Matem\'aticas, Grupo Interdisciplinar de Sistemas Complejos (GISC), Avenida de la Universidad, 30 (edificio Sabatini), 28911 Leganés (Madrid), Spain}
\author{Carlos P\'erez-Espigares}
\email[]{carlosperez@ugr.es}
\affiliation{Departamento de Electromagnetismo y F\'isica de la Materia, Universidad de Granada, Granada 18071, Spain}
\affiliation{Institute Carlos I for Theoretical and Computational Physics, Universidad de Granada, Granada 18071, Spain}

\date{\today}

\begin{abstract}
Due to the deterministic nature of chaotic systems, fluctuations in their trajectories arise solely from the choice of initial conditions. Some of these dynamical fluctuations may lead to extremely unlikely scenarios. Understanding the impact of such rare events and the trajectories that give rise to them is of significant interest across disciplines. Yet, identifying the initial conditions responsible for those events is a challenging task due to the inherent sensitivity to small perturbations of chaotic dynamics. In a recent paper [Phys. Rev. Lett. {\bf{131}}, 227201 (2023)], this challenge was addressed by finding the effective dynamics that make rare events typical for one-dimensional chaotic maps. Here we extend such large-deviation framework to $d$-dimensional chaotic maps. Specifically, for any such map, we propose a method to find an effective topologically-conjugate map which reproduces the rare-event statistics in the long-time limit. We demonstrate the applicability of this result using several observables of paradigmatic examples of two-dimensional chaos, namely the two-dimensional tent map and Arnold's cat map.

\end{abstract}

\maketitle



\section{Introduction}\label{sec:introduction}
The hallmark of chaotic systems is their extreme sensitivity to perturbations: even tiny differences in phase space can eventually result in dramatically different time evolutions \cite{ott02,strogatzbook}. In fact, due to their deterministic nature, fluctuations of trajectory-dependent observables in such systems can arise solely from variations in their initial conditions. It is frequently the case that, for long times, most trajectories produce typical values of those observables, with only a few leading to highly improbable outcomes. Such rare events can often be the most significant, however, as they may have a substantial positive or negative impact. This may be particularly important in fields such as climate science and financial markets, where the consequences of extreme events—e.g.,~natural disasters or financial crashes—can be severe \cite{ragone2018,Galfi2021,Klioutchnikov2017}. Hence, the widespread interest in understanding how such events arise, and, even more so, how they can be harnessed in practice to ensure that dynamical behaviors remain within acceptable bounds.

A natural approach to the study of rare events in chaotic systems is through large deviation theory, which provides a mathematical framework for quantifying the likelihood of rare fluctuations that decay exponentially in time \cite{touchette09}. By a statistical characterization of deviations from the typical values of time-averaged observables, this theory allows one to estimate the probability of rare events and gain insight into the mechanisms that trigger them. This has been amply illustrated in several stochastic systems of physical interest, including diffusive processes \cite{chetrite13a,chetrite15a,chetrite15b,angeletti16a,tsobgni16a}, lattice gas models \cite{simon2009,popkov10a,jack2010,marcantoni20a,hurtadogutierrez20,gutierrez21a,gutierrez21b,hurtadogutierrez23}, kinetically-constrained models of soft condensed matter  \cite{garrahan07a,garrahan09a,banuls19a,causer20a} and open quantum systems \cite{garrahan10a,carollo18b}.  Further, a large deviation approach has been also put forward to study dynamical fluctuations of Lyapunov exponents in different chaotic systems \cite{tailleur07,laffargue13a}. Yet, it is only recently that the range of cases amenable to such analysis has been extended so as to include general observables of deterministic dynamics of chaotic maps \cite{Smith22,Gutierrez2023,Monthus2024,Defaveri2025}, including also a random dynamical system lying at the interface between deterministic and random dynamics \cite{Monthus2025}. In the particularly important case of information-theory and fractal-geometry observables, however, the study of deterministic systems displaying chaos has employed large-deviation reasoning for decades \cite{ feigenbaum1989, beck93}.
\rev{In addition, large deviation methods have been  developed in the context of deterministic dynamical systems related to geophysical fluid flows, in order to study the likelihood of appearance of rogue waves \cite{vaneij18a} and tsunamis \cite{tong21a}, including a variational approach to identifying phase-space regions leading to extreme events in turbulent systems \cite{faraz17a}.} 

In fact, theoretical approaches more or less explicitly based on large deviations are common throughout statistical and nonlinear physics, including the foundations of equilibrium statistical mechanics itself \cite{touchette09, ellis84a,oono89a}. As in the classical theory of statistical-mechanical equilibrium ensembles, in order to study rare events, the probabilities are usually reweighed by an exponential tilting. In the present dynamical context, these are probability distributions of dynamical observables defined on trajectories, instead of thermodynamic observables defined on (static) configurations. Such biasing gives rise to a new trajectory ensemble known as the $s$-ensemble \cite{garrahan09a}, which is akin to the canonical (Gibbs) ensemble for fixed temperature and fluctuating energy, but in this case for fixed biasing field $s$ and a fluctuating time-integrated observable. 

While the biased ensemble encodes the statistics of the rare events of interest, sampling efficiently exponentially unlikely events from it is very challenging. \rev{Indeed, this issue lies at the basis of importance sampling, and the simulation of rare events more generally, an area of applied mathematics strongly related to the theory of large deviations; see \cite{Bucklew2004} for an introduction.}   In the context of stochastic systems, \rev{at least} two importance-sampling schemes have been put forward for this purpose: transition path sampling \cite{bolhuis02a,hedges09a} and the cloning method \cite{giardina06a,lecomte07a,giardina11a}. Although these approaches are often highly efficient, it has been shown that their performance can be further enhanced using trajectory umbrella sampling by means of an alternative sampling dynamics \cite{ray18a,klymko18a}. In fact, there exists an optimal sampling dynamics—known as the generalized Doob transform \cite{simon2009,popkov10a,jack2010,chetrite15b,carollo18b}—which directly yields the exponentially biased trajectory ensemble, effectively rendering rare events typical. Constructing this optimal dynamics requires solving a large-deviation problem, which amounts to a spectral analysis of an auxiliary biased (or ``tilted'') evolution operator, as the Doob transform depends on its largest eigenvalue and corresponding eigenvectors. This task is generally challenging, except for systems of moderate size. Thus, adaptive schemes have been successfully developed to approximate the optimal dynamics with iteration and feedback mechanism \cite{nemoto16a,ferre18a,das19a}, through variational tensor network approaches \cite{gorissen09a,banuls19a} or reinforcement learning techniques \cite{das21a,rose21a,gillman24a,pamulaparthy25a}. As a result, events that were extremely rare under the original dynamics become typical under the optimal one. This approach may be useful in order to modify a given dynamics for dynamical control purposes or just to gain insight into the nature of atypical trajectories in the original dynamics.

Despite its success in the stochastic realm, the application of the Doob transform in chaotic systems has only been proposed recently. In particular, it has been derived for general ergodic one-dimensional chaotic maps \cite{Gutierrez2023}, by relying on discretizations of continuous phase spaces and numerical approximations, and then more deeply studied via exact analytical techniques in connection with the doubling map \cite{Monthus2024}, which is particularly well suited for analytical treatment.  Notably, unlike the Doob transform in stochastic systems --which involves a reweighing of the original transition rates--, its counterpart for chaotic maps, due to their deterministic nature, entails a topological conjugation of the original map. This means that the unlikely trajectories underlying a rare event are generated with a suitably modified map that preserves many salient properties of trajectories (existence and stability of periodic orbits, Lyapunov exponents, ergodicity, etc.) of the original system.

In this work, we extend the derivation of the Doob transform proposed in Ref.~\cite{Gutierrez2023}, which crucially required systems to be one-dimensional, to the more general setting of $d$-dimensional chaotic maps. The methodology is illustrated by applying it to various observables of the two-dimensional tent map and Arnold's cat map. The paper is structured as follows. In Sec.~\ref{stat_traj}, the statistics of dynamical observables over trajectories are addressed, including concepts related to exponentially-tilted distributions and large deviations. In Sec.~\ref{doob_sec}, a method is proposed to obtain the generalized Doob transform of the chaotic dynamics and the ensuing topologically-conjugate map that produces the rare events of the original dynamics as typical statistics in its invariant measure. In Sec.~\ref{sec:applications}, the approach is applied to the study of the above-mentioned two-dimensional maps. Finally, in the concluding remarks, we summarize our methodological contribution and our main findings and propose some ideas for future developments in the study of large deviations of chaotic dynamics. \rev{More technical aspects of our work are treated in two appendices, which provide a discussion of numerical convergence issues, and a description of the computational procedure that we employ to arrive at the results discussed in Sec.~\ref{sec:applications}, respectively.}

\section{Statistics of trajectories in $d$-dimensional chaotic maps}
\label{stat_traj}

\subsection{Natural dynamics}
\label{nat}

We consider a discrete-time dynamical system whose evolution is given by
\begin{equation}
\vx_{n+1}=\vf(\vx_n),
\label{map}
\end{equation}
where $\vx_n \in I^d$ denotes the $d$-dimensional state vector at the $n$-th time step, for $n=0,1,2,...$, and the phase-space $I^d=I_1\times I_2\times ... I_d$ is the Cartesian product of $d$ compact intervals of the real line. The map $\vf: I^d\to I^d$ is assumed to generate a chaotic dynamics that is ergodic with respect to a natural invariant measure for almost every initial condition.

To study such a deterministic system from a statistical viewpoint, one considers the probability density of initial values, $\alpha_0(\vx)$, which evolves according to
\begin{equation}
\alpha_{n+1}(\vx)=L[\alpha_n(\vx)]
\label{probevol}
\end{equation}
where $L$ is the Frobenius-Perron operator (see, e.g., Ref.~\cite{beck93}), given by
\begin{equation}
L[\alpha(\vx)] = \int_{I^d} \alpha(\vy)\, \delta(\vx-\vf(\vy))\, d\vy.
\label{FP}
\end{equation}
This is an integral over probabilities restricted to the preimages of $\vx$ under $\vf$, as indicated by the Dirac delta $\delta(\vx-\vf(\vy))$. According to our assumptions, in the long-time limit the probability density converges to the natural invariant density of the map, denoted by $\rho(\vx)$, so that $\rho(\vx)=L[\rho(\vx)]$. Further, the adjoint or Koopman operator reads
\begin{equation}
L^\dagger[\alpha(\vx)]  = \alpha(\vf(\vx)),
\label{Koop}
\end{equation}
which verifies $\langle\alpha, L[\beta]\rangle = \langle L^\dagger[\alpha],\beta\rangle$ for the standard inner product, 
\begin{equation}
\langle\alpha, \beta\rangle = \int_{I^d}\alpha(\vx)\beta(\vx) d\vx\,.
\label{inner}
\end{equation}
It is worth noting that, for $\alpha(\vx)=\mathbbm{1}(\vx)$ (with $\mathbbm{1}(\vx)=1, \forall \vx\in I^d$), probability conservation, i.e., \! $\int L[\beta(\vx)] d\vx=\int \beta(\vx)d\vx=1$, implies that $L^\dagger[\mathbbm{1}(\vx)]=\mathbbm{1}(\vx)$ almost everywhere (a.e.), i.e.,~except perhaps in a set of Lebesgue measure zero. Hence, we conclude that the eigenfunctions of the Frobenius-Perron operator (\ref{FP}) and its adjoint (\ref{Koop}) associated with the eigenvalue having the largest real part, which is $1$, are the invariant density $\rho(\vx)$ and $\mathbbm{1}(\vx)$, respectively (or any members of the equivalence class of functions that are equal to them a.e.). 

\subsection{Time-integrated observables}

We focus on the statistics of a time-averaged observable $A_N$ across a trajectory of $N$ time-steps starting from the initial condition $\vx_0$, given by 
\begin{equation}
A_N(\vx_0)=\frac{1}{N}\sum_{n=0}^{N-1}g(\vx_n) = \frac{1}{N}\sum_{n=0}^{N-1}g(f^{n}(\vx_0)),
\label{obs}
\end{equation}
where the scalar $g(\vx)$, $g:I^d \to \mathbb{R}$, is the local contribution to the observable of interest. The second equality in Eq.~(\ref{obs}) is only meant to highlight that the value of the observable in a trajectory of $N$ steps is in fact determined by the initial condition $\vx_0$: $\vx_n = f^{n}(\vx_0)$, where $f^{n}$ is the composition of $f$ with itself $n$-times, and $f^0$ is the identity map. Since we assume that the map $f$ is ergodic with respect to its invariant measure given by the density $\rho(\vx)$, the long-time average of the observable converges to the ensemble average,
\begin{equation}
\la A \ra = \lim_{N\to \infty} A_N(\vx_0) =\int_{I^d}g(\vx)\rho(\vx)d\vx\, .
\label{ergod}
\end{equation}

We are interested in the statistics of $A_N$ \eqref{obs} across trajectories of $N$ steps, the probability of each trajectory $\vx_0 \to \vx_1\to \cdots \to \vx_{N-1}$ being given by the density of initial conditions $\alpha_0(\vx)$. In such ensemble of trajectories, the probability for the observable to take a value $a$ in a trajectory of $N$ steps is 
\begin{equation}
P_N(A_N=a) = \la \delta\left(A_N(\vx)-a\right) \ra_{\alpha_0} \sim e^{-NI(a)}
\label{LDP}
\end{equation}
where $\la\cdot  \ra_{\alpha_0} = \int_{I^d} \cdot\, \alpha_0(\vx)\, d\vx$. From now on, this probability density, which we simply denote as $P_N(a)$, is assumed to be dominated by the exponential form given above for $N\gg 1$. The validity of this assumption, which is frequently made (and leads to results consistent with it) in the analyses of rare events in stochastic systems, will be vindicated by our results below. This was also the case in the chaotic maps considered in Refs.~\cite{Smith22,Gutierrez2023, Monthus2024, Defaveri2025}, which suggests that the assumption holds quite generally. 

The large deviation function (LDF) \cite{touchette09} $I(a)$ in Eq.~\eqref{LDP}, also known as the rate function, gives the rate at which the probability concentrates around $\la A \ra = \lim_{N\to \infty} \int a\, P_N(a)\, da$ as $N$ increases. This function is positive, $I(a)\ge 0$, and is equal zero at $\la A \ra$, fluctuations away from it becoming exponentially unlikely in time. 
Dynamical large-deviation theory \cite{touchette09} indeed focuses on rare fluctuations whose probability [which becomes exponentially suppressed in time, cf.\! Eq.~(\ref{LDP})] is governed by the rate function $I(a)$, and the trajectories giving rise to them. In particular, finding a map that generates these rare trajectories is the core objective of this work.

\subsection{Biased ensemble of trajectories}

A direct estimation of the LDF $I(a)$ is typically very difficult, so an alternative way to find relevant information about the distribution $P_N(a)$ \eqref{LDP} is frequently adopted. It is based on the moment-generating function 
\be
Z_N(s)= \la e^{-s N A_N(\vx)} \ra_{\alpha_0} = \int   e^{-s N a}\, P_N(a)\, da \sim e^{N\theta(s)},
\label{zs1}
\ee
which also follows a large-deviation principle, in this case given by the scaled cumulant-generating function (SCGF) $\theta(s)$, whose derivatives are proportional to the cumulants of $A_N$ \cite{touchette09}. From \eqref{LDP} and \eqref{zs1}, applying a saddle-point approximation for long times $N\gg 1$, the SCGF arises as the Legendre transform of the LDF,
\be
\theta(s)=-\min_{a}[I(a)+sa].
\label{leg}
\ee
Indeed, $I(a)$ and $\theta(s)$ serve as thermodynamic potentials in the context of trajectory statistics, playing roles that are analogous to those of the entropy and the Helmholtz free energy in standard equilibrium statistical mechanics.  From this perspective, $Z_N(s)$ \eqref{zs1} is a dynamical partition function, which also corresponds to the normalization factor of the  exponentially-tilted distribution
\be
P_{N,s}(a)=\frac{e^{-s N a} P_N(a)}{Z_N(s)},
\label{tiltdist}
\ee
which biases the original distribution towards atypical fluctuations through the parameter $s$. 

The tilting of the distribution that results in $P_{N,s}(a)$ \eqref{tiltdist} thus amounts to a change from an (microcanonical) ensemble of trajectories   with a fixed value of $A$, defined by $P_N(a)$ \eqref{LDP}, to a new (canonical) ensemble of trajectories ---the $s$-ensemble \eqref{tiltdist}---, where $A_N$ fluctuates across trajectories \cite{garrahan09a}, yet we fix the average, 
\be
\la A \ra_s = \lim_{N\to \infty} \int a\, P_{N,s}(a)\, da ,
\label{ave_s}
\ee
by tuning the value of $s$. In terms of probabilities over trajectories of $N$ steps, ``flat'' averages over initial conditions $\la \cdot \ra_{\alpha_0}$ are replaced by exponentially-weighted averages $\la \cdot \ra_s=\la \cdot \, e^{-s N A_N(\vx)}\ra_{\alpha_0}/Z_N(s)$. Following the analogy with equilibrium statistical mechanics, $s$ here would act as an inverse temperature, with the difference that it can be either positive (which favors values of $A$ smaller than the unbiased average $\la A\ra_{s=0} = \la A \ra$) or negative (which favors values greater than $\la A\ra$). For a given fluctuation $a$, the appropriate choice of $s$ is the one that yields it as the average value, i.e.,\! $\la A \ra_s = a$. In terms of the SCGF and the LDF, this corresponds to $-\theta'(s)=a$ and $I'(a)=-s$, respectively \cite{touchette09}. 

\subsection{Tilted Frobenius-Perron operator}
\label{subsec:tilted}

This subsection is based on the analogous reasoning developed in Ref.~\cite{Gutierrez2023} for one-dimensional maps, where each step is justified. The dynamical partition function $Z_N(s)$, Eq.~\eqref{zs1}, can be expressed as
\be
Z_N(s)=\int_{I^d}L_s^N[\alpha_0(\vx)]d\vx,
\label{zs2}
\ee
where $L_s$ is the tilted Frobenius-Perron operator \cite{Smith22}, 
\be
L_s[\alpha(\vx)]=\int_{I^d} e^{-s g(\vy)}\alpha(\vy) \delta(\vx-\vf(\vy))\, d\vy.
\label{tiltop}
\ee
It is then straightforward to check, by focusing on the spectrum of $L_s$, that for long times $Z_N(s)$ follows the large deviation form $Z_N(s)\sim e^{N\theta(s)}$ [in agreement with Eq.~\eqref{zs1}] where the exponential of the SCGF \eqref{leg}, $e^{\theta(s)}$, must be the eigenvalue having the largest real part of $L_s$ and its adjoint, 
\bea
L_s[r_s(\vx)] &=& e^{\theta(s)} r_s(\vx),\nonumber \\
L^{\dagger}_s[l_s(\vx)] &=& e^{\theta(s)} l_s(\vx).
\label{eigprob}
\eea
Here $r_s(\vx)$ and $l_s(\vx)$ are the right and left eigenfunctions, respectively, which are normalized so that $\int_{I^d} r_s(\vx)l_s(\vx)d\vx=1$ and $\int_{I^d} r_s(\vx)d\vx=1$. And the adjoint of the tilted Frobenius-Perron operator is given by
\be
L^{\dagger}_s[\alpha(\vx)]=e^{-s g(\vx)} \alpha[\vf(\vx)].
\label{tiltadjop}
\ee
The sampling problem of finding the LDF $I(a)$ is thus transformed into an eigenvalue problem, which involves determining the eigenvalue with the largest real part 
as a function of $s$ \eqref{eigprob}. From such an eigenvalue, we derive the SCGF $\theta(s)$, which then allows us to calculate $I(a)$ through the inverse of the Legendre transformation \eqref{leg}.

\rev{A few comments are in order regarding the practical implementation of these steps, which is based on numerically solving a discretized version of the eigenproblem in Eqs.~(\ref{eigprob}). For that purpose, the tilted Frobenius-Perron operator $L_s$ and its adjoint $L_s^\dagger$ are approximated by matrices, and the eigenfunctions $r_s(\vx)$ and $\l_s(\vx)$ are approximated by the eigenvectors of those matrices. The discretization of Frobenius-Perron operators, resulting in stochastic matrices (see e.g.~Ref.~\cite{li1977}), and variations thereof (see e.g.~the case of Ruelle-Frobenius-Perron operators in Ref.~\cite{Froyland07}, of which tilted Frobenius-Perron operators are a particular case) has a long-standing history, and is sometimes referred to as Ulam's method. Seemingly, it first appeared in connection with a conjecture by S. Ulam to the effect that the right eigenvector of the discretized operator, for which the Frobenius-Perron theorem of positive matrices may be applied, should converge to the invariant density of the original (infinite-dimensional) Frobenius-Perron operator; see \cite{UlamMethod}, where the author mentions computational evidence based on the logistic map. (This conjecture has been proven since then for some families of maps, see e.g.~Ref.~\cite{li1977}.) Likewise, in our computations, Eq.~(\ref{zs2}) is taken to be a discretized version of this same equation, yielding analogous expressions to those commonly employed for (finite-dimensional) tilted operators in the study of large deviations of Markov chains (see e.g.~\cite{gutierrez2021}). On the other hand, exact versions of equations similar to Eq.~(\ref{zs2}) involving characteristic functions have been rigorously studied in the mathematical literature on ergodic theory (see e.g.~\cite{Hennion2001} and references therein).}

\subsection{Statistical characterization of rare events}\label{subsec:statistical}

If the SCGF $\theta(s)$ is known, one can adjust the mean of the distribution to a rare fluctuation, $\la A \ra_s=a$, by choosing $s$ so that $-\theta'(s)=a$ or, equivalently, $-s=I'(a)$. The focus of this work lies in the trajectories themselves that sustain such a fluctuation. More specifically, our aim is to find a map which generates the trajectories reproducing the statistics of the $s$-ensemble, having a long-time average of the observable equal to $\la A \ra_s$. This is analogous to \eqref{obs} for $N\to \infty$, see also \eqref{ergod},  but for a new dynamical map instead of $\vf$, the latter being recovered in the unbiased case $s=0$. 

To this end, we consider the time average of an observable $B_N$, in general different from the one used in the biasing $A_N$, defined over a trajectory of $N$ steps starting from $\vx_0$,
\be
B_N(\vx_0)=\frac{1}{N}\sum_{n=0}^{N-1} k(\vx_n).
\label{obsB}
\ee 
The $s$-ensemble average reads
\be
\la B_N \ra_s = \frac{\la B_N(\vx)\,   e^{-s N A_N(\vx)} \ra_{\alpha_0}}{Z_N(s)},
\ee
and in the following analysis we apply insights derived from the large deviations of jump processes \cite{garrahan09a}. In that context, it is known that a large-deviation event suffers from time-boundary effects; therefore, it must be computed as an average in the time-bulk of the trajectory. In particular, $\la B_N \ra_s$ must be equal to the average of the local contribution $k(\vx_\nu)$ at a time step $\nu$  that lies in the `middle' of the trajectory, i.e., ~$0\ll \nu \ll N$. Hence,
\be
\la B_N \ra_s=\frac{\la k(\vf^\nu (\vx))\, e^{-s N A_N(\vx)} \ra_{\alpha_0}}{Z_N(s)}.
\label{midstat}
\ee

To compute this average, we shift to operatorial (Dirac) notation by introducing an orthonormal basis of phase-space points, $\{\ket{\vx}\}$, satisfying $\braket{\vx}{\vy}=\delta(\vx-\vy)$. The state of the system is given by its probability distribution at time $n$, $\alpha_n(\vx)$, which is encoded in the ket $\ket{\alpha_n}=\int_{I^d} \alpha_n(\vx) \ket{\vx} d\vx$, with $\alpha_n(\vx)=\braket{\vx}{\alpha_n}$. The Frobenius-Perron and the tilted operator act as, $L[\alpha_{n}(\vx)]=L\ket{\alpha_n}$ and $L_s[\alpha_{n}(\vx)]=L_s\ket{\alpha_n}$, respectively. We further introduce the flat state, $\bra{-}=\int_{I^d}\bra{\vx}d\vx$, such that conservation of probability is conveniently expressed as $\braket{-}{\alpha_n}=1$. The observable $k(\vx)$ can then be written as a diagonal operator $\hat{k}=\int_{I^d} k(\vx) \ketbra{\vx}{\vx} d\vx$ with $k(\vx)=\bra{\vx}\hat{k}\ket{\vx}$. With this notation, the dynamical partition sum \eqref{zs2} reads $Z_N(s)=\la e^{-s N A_N(\vx)} \ra_{\alpha_0}=\bra{-}L_s^N\ket{\alpha_0}$. And the average $\la B_N \ra_s$ in the bulk of the trajectory, given by \eqref{midstat}, is now 
\be
\la B_N \ra_s =\frac{\bra{-}L_s^{N-\nu}\hat{k}L_s^{\nu}\ket{\alpha_0}}{\bra{-}L_s^N\ket{\alpha_0}}.
\label{midstat2}
\ee
Based on the spectral decomposition of $L_s=e^{\theta(s)}\ketbra{r_s}{l_s}+\cdots$, for $0\ll \nu \ll N$, the terms in the numerator become
\be
L_s^{\nu}\ket{\alpha_0}\approx e^{\nu\theta(s)}\ket{r_s}\braket{l_s}{\alpha_0}
\label{spec1}
\ee
and
\be
\bra{-}L_s^{N-\nu}\approx e^{(N-\nu)\theta(s)}\braket{-}{r_s}\bra{l_s}
\label{spec2}
\ee
and the denominator can be approximated as,
\be
\bra{-}L_s^N\ket{\alpha_0}\approx e^{N\theta(s)}\braket{-}{r_s}\braket{l_s}{\alpha_0}.
\label{spec3}
\ee
Putting Eqs.~\eqref{spec1}, \eqref{spec2} and \eqref{spec3} into the average in Eq.~\eqref{midstat2} and simplifying, we readily check that (in the limit $0\ll \nu \ll N$)
\be
\la B_N \ra_s=\bra{l_s}\hat{k}\ket{r_s} = \int_{I^d}k(\vx)l_s(\vx)r_s(\vx)d\vx.
\label{midstat3}
\ee

\rev{The spectral decomposition right above Eq.~(\ref{spec1}), on which Eq.~(\ref{midstat3})  is based, assumes the existence of a non-degenerate largest eigenvalue. (Degenerate or nearly degenerate spectra would lead to dynamical phase transitions and metastability, respectively; see Refs.~\cite{hurtadogutierrez23} for examples in the context of stochastic dynamics.) Again, the decomposition implies that the tilted Frobenius-Perron operator $L_s$ has been discretized, hence approximated by a positive matrix on which the Frobenius-Perron theorem should be applicable. In a rigorous infinite-dimensional setting, spectral gaps may be connected to the quasi-compacity of Frobenius-Perron operators, which allows for the description of the long-term behavior of densities in terms of one or finitely many eigenfunctions (see e.g.~Ref.~\cite{Hennion2001} and references therein).}

Hence, \rev{from Eq.~(\ref{midstat3})}, analogously to what is found for jump processes in Ref.~\cite{garrahan09a}, we conclude that trajectories corresponding to the $s$-ensemble distribution, however they are generated, must have as natural invariant density
\be
\rho_s(\vx)=l_s(\vx)r_s(\vx),
\label{sinv}
\ee
which of course reduces to $\rho(\vx)$ for the unbiased dynamics $s=0$.
The task now is to derive a Frobenius-Perron operator having such an invariant density and, ultimately, to propose an effective map corresponding to it. This will lead us to the main results of this work, namely to the generalized Doob transform for $d$-dimensional chaotic maps and the derivation of the effective ($d$-dimensional) map naturally generating the rare trajectories of interest as typical events, as we shall show in the next section.

\section{Doob effective dynamics in $d$-dimensional chaotic maps: How to make rare events typical}
\label{doob_sec}

\subsection{Generalized Doob transform}
\label{doobtrans}

In this section, we assume that a particular value of the tilting parameter is chosen, $s= s_0$. Such $s_0$ is the value that results in the statistics of interest of the observable under consideration in the $s$-ensemble, as given by the density $\rho_{s_0}(\vx)=l_{s_0}(\vx)r_{s_0}(\vx)$. Following Ref.~\cite{Gutierrez2023}, we introduce the so-called Doob generator, which for $d$-dimensional maps generalizes in a straightforward manner to 
\begin{equation}
L_{s_0}^D[\alpha(\vx)] = e^{-\theta(s_0)} l_{s_0}(\vx) L_{s_0}[\left(l_{s_0}(\vx)\right)^{-1}\alpha(\vx)].
\label{LD}
\end{equation}
This is a Frobenius-Perron operator that satisfies the desired conditions: (i) it is a proper probability-preserving generator, $(L_{s_0}^D)^{\dagger}[\mathbbm{1}(\vx)]=\mathbbm{1}(\vx)$, which the tilted operator \eqref{tiltop} is not, \rev{(ii) its SCGF is $\theta^D(s) = \theta(s_0+s) - \theta(s_0)$, hence its derivatives of order $n$ satisfy $(\theta^D)^{(n)}(0) = \theta^{(n)}(s_0)$, and (iii)} its invariant density is $\rho_{s_0}(\vx)$, $L_{s_0}^D[\rho_{s_0}(\vx)] = \rho_{s_0}(\vx)$ \footnote{As in Subsec.~\ref{nat} we do not distinguish between densities that are equal a.e.}. \rev{Property (ii) implies that the cumulants of the new dynamics are those of the rare fluctuation for $s=s_0$ in the original dynamics, hence $L_{s_0}^D$ in Eq.~(\ref{LD})} generates the ensemble of trajectories yielding the tilted distribution \eqref{tiltdist} in the long-time limit, as shown in detail in the Supplemental Material of Ref.~\cite{Gutierrez2023}.

The Doob generator \eqref{LD}, however, is an operator that acts on probability densities $\alpha(\vx)$, just as any other Frobenius-Perron operator. Yet we are ultimately interested in finding a deterministic map acting on state vectors $\vx$ that, starting from an arbitrary initial condition, typically generates the trajectories that sustain the rare events corresponding to $s=s_0 \neq 0$. In this regard, while obtaining the Frobenius-Perron operator for a given map $\vf$ is, at least formally, easy [see Eq.~\eqref{FP}], it is not obvious how to find a map that corresponds to a given Frobenius-Perron operator.  What follows is a procedure that results in a chaotic map, referred to as the Doob effective map, different from the original one $\vf$, which \rev{naturally generates the rare flucutations corresponding to some $s = s_0$}. It is based on a transformation $\vxt=\gamma_{s_0}(\vx)$ of the state vectors $\vx$ taken over (typical) trajectories of $\vf$ --which yield a distribution $\rho(\vx)$ for long times, see Eq.~\eqref{ergod}-- with the right properties so that the long-time distribution of the new coordinates $\vxt$ is given by  $\rho_{s_0}(\vxt)$.

\subsection{Doob effective map in $d \geq 1$ dimensions}
\label{doobgen}

The following subsections focus on how to find an explicit expression for the transformation $\vxt = \vg_{s_0}(\vx)$, where $\vx, \vxt \in I^d \subset \mathbb{R}^d$, first for the one-dimensional case, $d=1$ (this will be a review of results from Ref.~\cite{Gutierrez2023}), and then their generalization to $d$-dimensions, $d>1$. But before presenting the detailed derivation of the transformation $\vg_{s_0}(\vx)$, we here illustrate the way in which the Doob effective map can be built from it, and discuss why it achieves the expected result and some of its properties. The discussion is relevant for any number of dimensions.

From the original map $\vf$ giving the evolution $\vx_{n+1}=\vf(\vx_n)$, we can easily derive the evolution for $\vxt$, $\vxt_{n+1}=\vf^D_{s_0}(\vxt_n)$, i.e.~the explicit expression for the Doob effective map $\vf^D_{s_0}$, assuming knowledge of the (invertible) transformation $\vg_{s_0}(\vx)$. Since $\vxt_{n+1}=\vf^D_{s_0}(\vxt_n)=\vf^D_{s_0}[\vg_{s_0}(\vx_n)]$, and also $\vxt_{n+1}=\vg_{s_0}(\vx_{n+1})=\vg_{s_0}[\vf(\vx_n)]$, we obtain that $\vf^D_{s_0}[\vg_{s_0}(\vx_n)]=\vg_{s_0}(\vf(\vx_n))$, then  $\vf^D_{s_0}(\vxt_n)=\vg_{s_0}\{\vf[\vg_{s_0}^{-1}(\vxt_n)]\}$. Hence, the Doob effective map will be defined as follows,
\be
\vf^D_{s_0}=\vg_{s_0} \circ \vf\circ \vg_{s_0}^{-1},
\label{fDDd}
\ee 
a relation that is illustrated in the following diagram
\begin{center}
\begin{tikzcd}
{\vx}_n \arrow[r, "\vf"]
& {\vx}_{n+1} \arrow[d, "\vg_{s_0}" ] \\
\vxt_n \arrow[r, "\vf^D_{s_0}"]\arrow[u, "\vg_{s_0}^{-1}"]
&  \vxt_{n+1}
\end{tikzcd}
\end{center}
Equivalently, $\vf = \vg_{s_0}^{-1} \circ \vf^D_{s_0}\circ \vg_{s_0}$.

\begin{figure}
    \centering
    \includegraphics[width=1\linewidth]{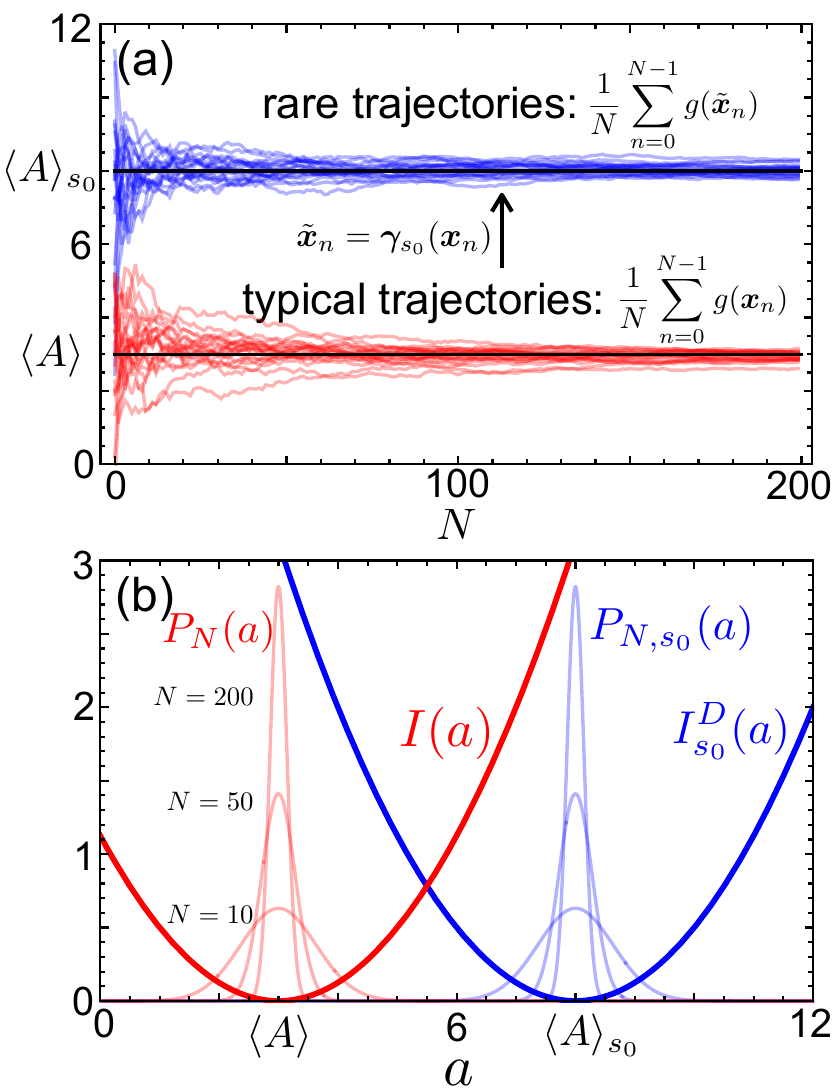}
    \caption{Sketch of the main idea behind making rare events typical by a topological conjugation \eqref{fDDd}. (a) The red lines show the original trajectories $\vx_0 \to \vx_1 \to \vx_2 \to \cdots \to \vx_{N-1}$ as reflected in the value of the observable $A_N = N^{-1} \sum_{n=0}^{N-1} g(\vx_{n})$, converging to the long-time limit $\la A\ra$ \eqref{ergod}. The blue lines show trajectories of the transformed dynamics $\vxt_0 \to \vxt_1 \to \vxt_2 \to \cdots \to \vxt_{N-1}$, where $\vxt_n = \gamma_{s_0}(\vx_n)$, as reflected in the value of the same observable $N^{-1} \sum_{n=0}^{N-1} g(\vxt_n)$ evaluated at the transformed points, converging to the $s$-ensemble average $\la A \ra_{s_0}$ \eqref{ave_s2}. (b) The probability distribution of  $A_N = N^{-1} \sum_{n=0}^{N-1} g(\vx_n)$, $P_N(a)$ \eqref{LDP}, concentrates around $\la A \ra$ for large $N$, with LDF $I(a)$, while the probability distribution of the same observable in the transformed trajectory $\vxt_n = \gamma_{s_0}(\vx_n)$ concentrates around $\la A \ra_{s_0}$ for large $N$, which is the first moment of $P_{N,s_0}(a)$ \eqref{tiltdist}, with LDF $I^D_{s_0}(a)$.}
    \label{fig1}
\end{figure}

It is interesting to observe that for chaotic maps the Doob effective map is topologically conjugate to the original dynamics \eqref{fDDd}, in contrast to the stochastic case where the Doob dynamics corresponds to a reweighing of the original (unbiased) transition probabilities between configurations \cite{jack2010,chetrite15b}.  Because topological conjugation preserves ergodicity, the long-time average of the observable in the transformed (rare) trajectories converges to its $s$-ensemble average for $s=s_0$ [i.e.,~the average taken with respect to the invariant density $\rho_{s_0}(\vxt)$],
\be
\la A \ra_{s_0} =\lim_{N\to \infty}\frac{1}{N}\sum_{n=0}^{N-1}g(\vxt_n)=\int_{I^d}g(\vxt)\rho_{s_0}(\vxt)d\vxt\, ,
\label{ave_s2}
\ee
cf.~Eq.~\eqref{ergod}. The main idea behind all this is illustrated in the sketch displayed in Fig.~\ref{fig1} (a), where the original (typical) trajectories are transformed according to $\vxt_n=\vg_{s_0}(\vx_n)$, giving rise to the $s$-ensemble of rare trajectories for $s=s_0$. The long-time average of the observable evaluated in such rare trajectories yields the rare even of interest, $\la A\ra_{s_0}$, corresponding to the tilted distribution $P_{N,s_0}(a)$ in Eq.~\eqref{tiltdist}, 
as depicted in Fig.~\ref{fig1}(b), which for long times takes the large-deviation form $P_{N,s_0}(a)\sim e^{-N I^D_{s_0}(a)}$ with $I^D_{s_0}(a)=s_0 a+I(a)+\theta(s_0)$. This rate function can be readily obtained by taking the long-time limit of \eqref{tiltdist} and considering that $P_N(a)\sim e^{-NI(a)}$ and $Z_N(s_0)\sim e^{N\theta(s_0)}$. In practice, as discussed above, $\theta(s_0)$ is obtained from the spectrum of the tilted generator \eqref{tiltop} and $I(a)$ can be extracted via its inverse Legendre transform, i.e.,\! the inverse of Eq.~\eqref{leg}. 

Topological conjugation being an equivalence relation, one might also inquire about its transitivity. In fact, applying the Doob conjugacy for $s = s_0$ followed by another conjugacy for $s= s_1$ results in a conjugacy for $s= s_0+s_1$, as seen by how multiple exponential tiltings combine in Eq.~(\ref{tiltdist}). In terms of the coordinate transformations, $\vg_{s_0}(\vx) \vg_{s_1}(\vx) = \vg_{s_0+s_1}(\vx)$, and in particular $\vg_{s_0}(\vx) \vg_{-s_0}(\vx) = \vg_{(s_0-s_0) = 0}(\vx) = \vx$ ($\vf_0^D = \vf$), which shows that $\vg_{s_0}^{-1}(\vx) = \vg_{-s_0}(\vx)$. Hence, the conjugacy resulting in the Doob effective map (\ref{fDDd}) is given by a one-parameter group of transformations with $s$ as parameter.

\subsection{Coordinate transformation in one dimension ($d=1$)}
\label{1dim}

Before embarking on a discussion of the topological conjugation in the $d$-dimensional case for $d>1$, we briefly review the solution for one-dimensional maps, $d=1$ \cite{Gutierrez2023}. In such a setting, finding $\xt=\gamma_{s_0}(x)$ is analogous to transforming values \( x \) of a random variable \( X \), which is distributed according to \( \rho(x) \), into values \( \xt \) of a random variable \( \xtm \), which follows the distribution \( \rho_{s_0}(\xt) \), by applying the inverse transformation method \cite{devroye06a}. In the original dynamics $x \in I\subset \mathbb{R}$, where $I$ is some compact interval (i.e.,~the phase space of the map $f$), and it turns out that we also have $\xt \in I$ by construction, since its density $\rho_{s_0}(\xt)$ is obtained from the eigenfunctions in Eq.~(\ref{eigprob}). In fact, $\gamma_{s_0}$ is a bijection on $I$, which we take to be increasing as in \cite{Gutierrez2023} (although the procedure that we are about to describe can be alternatively carried out for a decreasing transformation). 

Assuming that $\rho(x)$ and $\rho_{s_0}(\xt)$ are integrable and strictly positive \footnote{While the original invariant density $\rho(x)$ satisfies these properties in all cases that we consider, in the case of $\rho_{s_0}(\xt)$ (which is obtained from the eigenfunctions of the tilted Frobenius-Perron operator) the exact solution may display singular behavior \cite{Monthus2024}. This does not seem to affect our results in any serious way, given their approximate nature due to the numerical procedure based on phase-space discretization and the power method, yet it might be relevant when considering exact results based on maps amenable to analytical solution.}, their cumulative distributions $F(x) = \int^x \rho(u) du$ and $F^D_{s_0}(\xt) = \int^{\xt} \rho_{s_0}(u) du$ are continuous and increasing (hence invertible) \footnote{Whenever we omit the lower (upper) limit in one-dimensional integrals, they are integrated from the minimum (maximum) of their range, e.g.,~if a function is defined in $[a,b]$, then $\int^x$ will be an integration over $[a,x]$.}, then we can write
\be
\begin{aligned}
F^D_{s_0}(\xt)=P(\xtm\le \xt)=P(\gamma_{s_0}(X)\le \xt)=&\\
=P(X\le \gamma_{s_0}^{-1}(\xt))=&F[\gamma_{s_0}^{-1}(\xt)].
\end{aligned}
\ee
Therefore $F^D_{s_0}(\xt)=F(x)$ implies that 
\be
\xt=\gamma_{s_0}(x) = (F^D_{s_0})^{-1}[F(x)]. 
\label{g1d}
\ee
This expression yields the desired transformation. It is worth highlighting, as it will help establish connections with the $d$-dimensional case in Subsec.~\ref{ddim}, that the random variable defined by $Z =F^D_{s_0}(\tilde{X}) = F(X)$ is distributed uniformly as $U([0,1])$: $P(Z \leq z) = P(F(X) \leq z) = P(X \leq F^{-1}(z)) = F(F^{-1}(z)) = z$.

A generalization of this procedure is required for finding the Doob effective map in $d>1$ dimensions, Eq.~\eqref{fDDd}. This amounts to deriving a bijection $\vxt=\vg_{s_0}(\vx)$, with $\vx, \vxt \in I^d$, yielding new (transformed) trajectories $\vxt_0 \to \vxt_1 \to \vxt_2 \to \cdots \to \vxt_{N-1}$, that are distributed according to $\rho_{s_0}(\vxt)$ in the long-time limit. The mapping that is needed for the purpose was derived in a more general context as a solution for the inverse Frobenius-Perron problem (i.e.,~the problem of finding a map with a given invariant density) in Ref.~\cite{Fox2021}. We are going to employ the same approach here, which is based on the  Rosenblatt transform described below~\cite{Rosenblatt1952}, and is frequently applied to generate new
distributions in the so-called conditional distribution method~\cite{Dolgov2020}.

\subsection{Coordinate transformation in $d>1$ dimensions: Rosenblatt transform}
\label{ddim}

The Rosenblatt transform allows for the decomposition of the $d$-dimensional problem into the solution of $d$
one-dimensional transforms similar to those in Eq.~\eqref{g1d} using the notion of conditional probabilities. We shall closely follow the derivation of Refs.~\cite{Rosenblatt1952, Fox2021}. The probability distribution of interest depends on $d$ variables, and is absolutely continuous with respect to the density $\rho(\vx)=\rho(x_1,\dots,x_d)$, which is the invariant density of the original map $\vf$. This density can be expressed in the following form:
\begin{equation}
    \rho(x_1,\dots, x_d)=\rho_1(x_1)\rho_2(x_2|x_1)\cdots\rho_d(x_d|x_1,\dots,x_{d-1}),
\end{equation}
with the usual definition for conditional probabilities,
\begin{align}
    \rho_k(x_k|x_1,\dots,x_{k-1}) =\frac{\rho_k(x_1, \dots,x_{k-1},x_k)}{\rho_{k-1}(x_1, \dots, x_{k-1})},
    \label{eq:conditionals}
\end{align}
and of marginal joint probabilities,
\begin{equation}
    \rho_k(x_1, \dots,x_k)= \!\!\int_{I_{k+1}\times\dots\times I_{d}} \!\rho(x_1, \dots, x_d)\, dx_{k+1} \cdots dx_{d},
    \label{joint}
\end{equation}
where $k=1,\dots,d$, the joint probability for $k=d$ being just the invariant density, $\rho_d(x_1\dots,x_d) = \rho(x_1,\dots, x_d)$. In Eq.~\eqref{joint} it is the last $d-k$ variables that are integrated out, as indicated by the domain of integration.  

Rosenblatt introduced a transformation $(Z_1,\dots, Z_d) = T(X_1,\dots,X_d)$ in terms of cumulative probability distributions, such that the new set of variables are independent and uniformly distributed in the $d$-dimensional hypercube, $(Z_1,\dots, Z_d) \sim U([0,1]^d)$ \cite{Rosenblatt1952}. Such new variables $Z_1,\ldots, Z_d$ take values $z_1, \ldots, z_d$, which are defined in terms of those taken by $X_1,\ldots, X_d$, in turn denoted as $x_1, \ldots, x_d$, as follows:
\begin{align}
    z_1 = F_{1} (x_1) &= \int^{x_1} \rho_1(x_1')\,dx_1',\nonumber \\
z_2 = F_{2} (x_2|x_1) &= \int^{x_2} \rho_2(x_2'|x_1)\,dx_2',\nonumber \\
    &\vdots \nonumber\\
z_d = F_d(x_d|x_1,\dots,x_{d-1}) &=\int^{x_d} \rho_d(x_d'|x_1,\dots,x_{d-1})\, dx_d'.
\label{zs}
\end{align}
 Here, $F_k(x_k|x_1,\dots,x_{k-1})=P(X_k\le x_k|X_1=x_1,\dots,X_{k-1}=x_{k-1})$ for $k=1,\ldots, d$ is the conditional cumulative distribution, namely the cumulative distribution of $X_k$ having fixed the previous $k-1$ variables. 
 
Proceeding in the same way for points $\vxt=\vg_{s_0}(\vx)$, distributed following the invariant density of the tilted dynamics for $s=s_0$, $\rho_{s_0}(\vxt)=\rho_{s_0}(\tilde{x}_1,\dots,\tilde{x}_d)$, and applying again the Rosenblatt transformation, we obtain $(z_1,\dots, z_d) = T(\tilde{x}_1,\dots,\tilde{x}_d)$. The transformed variables $(Z_1,\dots, Z_d)$ are in fact the same that we obtained above for $(X_1,\dots,X_d)$ [distributed with density $\rho({\bm x})$], as given in Eqs.~\eqref{zs}, even if we are now applying the Rosenblatt transformation to $(\tilde{X}_1,\dots,\tilde{X}_d)$. The reason for this is that we assume that each variable $X_k$ can be transformed into $\tilde{X}_k$, for $k=1,\ldots, d$, by an increasing function $\vg_{s_0}$, an assumption that will be justified a posteriori below. Hence the equality of the cumulative distributions based on the original invariant density $\rho({\vx})$ and the Doob invariant density $\rho_{s_0}(\vxt=\vg_{s_0}(\vx))$, respectively: $\int^{x_1}\rho(x_1')\, d x_1' = \int^{\xt_1}\rho_{s_0}(\xt_1')\, d \xt_1'$, and the same applies to the other variables with more complex expressions based on conditional densities shown in Eq.~\eqref{zs}. Thus, the values taken by both sets of random variables are related as follows:
\begin{align}
   F_{1} (x_1) & = F_{s_0,1}^D (\tilde{x}_1),\nonumber \\
 F_{2} (x_2|x_1) & = F_{s_0,2}^D (\tilde{x}_2|\tilde{x}_1), \nonumber\\
 & \vdots\nonumber\\
 F_{k}(x_k|x_1,\dots,x_{k-1}) & = F_{s_0,k}^D(\tilde{x}_k|\tilde{x}_1,\dots,\tilde{x}_{k-1}),
\end{align}
where $F_{s_0, k}^D $ for $k=1,\ldots, d$ are the conditional cumulative distributions, each defined as the corresponding $F_k$ but with density $\rho_{s_0}$ instead of $\rho$.  By inverting these relations, we can obtain our conjugacy $\vg_{s_0}$. Introducing the informal notation $(F_{s_0,k}^D)^{-1}$ as the inverse of the cumulative distribution of $\tilde{X}_k$ given some fixed values $\tilde{X}_1 = \tilde{x}_{1},\dots, \tilde{X}_{k-1} = \tilde{x}_{k-1}$, we thus obtain an expression for each of the components of the transformation $\vxt=\vg_{s_0}(\vx)$:
\begin{eqnarray}
   \tilde{x}_1 =\gamma_{s_0,1}(x_1) & = &(F_{s_0,1}^D)^{-1}[F_{1} (x_1)], \label{gDd}\\ 
   \nonumber
   \tilde{x}_2 =\gamma_{s_0,2}(x_1,x_2)& =&(F_{s_0,2}^D)^{-1}[F_{2} (x_2|x_1)], \\ 
 & \vdots&\nonumber\\
 \tilde{x}_k =\gamma_{s_0,k}(x_1,\dots,x_k)& =&(F_{s_0,k}^D)^{-1}[F_{k} (x_k|x_1.\dots,x_{k-1})], 
 \nonumber
\end{eqnarray}
which is the $d$-dimensional extension of the method employed for the one-dimensional case as discussed in the previous subsection. In fact, in that setting, only the first equation in \eqref{gDd} is relevant, which is equivalent to Eq.~\eqref{g1d}.

Once the transformation $\vxt=\vg_{s_0}(\vx)$ given by \eqref{gDd} is known, one can readily obtain the Doob effective map in $d$-dimensions by applying the topological conjugation to the original map, Eq.~\eqref{fDDd}. 
Note that this map is not uniquely defined, as various equivalent forms of the transformation of coordinates $\vxt=\vg_{s_0}(\vx)$ may be used. Already within our methodology, apart from the choice of an increasing (instead of decreasing) function mentioned above, there is the fact that $d!$ different transformations can be derived, each corresponding to a different permutation of the variables $x_1,x_2, \ldots, x_d$ in Eqs.~\eqref{gDd}.

\subsection{Doob effective map in $2$ dimensions}

In the remainder of the paper, we shall put the procedure described above into practice in order to make rare events typical for several observables of two-dimensional maps displaying chaotic behavior, specifically the two-dimensional tent map and Arnold's cat map. Before doing that, we here explicitly derive the general expressions corresponding to the Doob effective map for $d=2$, since this is a particularly important case of the theory that will be at the basis of those numerical results displayed in the next section.

The transformation $(\tilde{x}_1,\tilde{x}_2)=\vg_{s_0}(x_1,x_2)=(\gamma_{s_0,1}(x_1),\gamma_{s_0,2}(x_1,x_2))$ can be obtained from the first and second lines in Eqs.~\eqref{gDd}.
We first compute 
\be
F_1(x_1) = \int^{x_1} \rho_1(x_1')\, dx_1'
\label{f01D}
\ee
with $\rho_1(x_1) = \int_{I_2} \rho(x_1, x_2')\, dx_2'$, where $\rho(x_1, x_2)$ is the invariant density of the original map. Further, 
\be
F_{2} (x_2|x_1) = \int^{x_2} \rho_2(x_2'|x_1)\, dx_2'
\label{f02D}
\ee
with $\rho_2(x_2|x_1)=\rho(x_1, x_2)/\rho_1(x_1)$. On the other hand, 
\be
F_{s_0,1}^D(\tilde{x}_1) = \int^{\tilde{x}_1} \rho_{s_0,1}(x_1')\, dx_1'
\label{fs01D}
\ee
with $\rho_{s_0,1}(\tilde{x}_1) = \int_{I_2} \rho_{s_0}(\tilde{x}_1, x_2')\, dx_2'$
and $ \rho_{s_0}(\tilde{x}_1, \tilde{x}_2)$ being the target invariant density. We can then calculate $\tilde{x}_1 =\gamma_{s_0,1}(x_1) = (F_{s_0,1}^D)^{-1}[F_{1} (x_1)]$. Now, we replace it in $F_{s_0,2}^D (\tilde{x}_2|\tilde{x}_1)$ to compute 
\be
F_{s_0,2}^D (\tilde{x}_2|\tilde{x}_1) = \int^{\tilde{x}_2} \rho_{s_0,2}(x_2'|\tilde{x}_1)\, dx_2'
\label{fs02D}
\ee
with $\rho_{s_0,2}(\tilde{x}_2|\tilde{x}_1)=\rho_{s_0}(\tilde{x}_1, \tilde{x}_2)/\rho_{s_0,1}(\tilde{x}_1)$. To conclude, we invert $F_{s_0,2}^D$ and obtain $\tilde{x}_2 =\gamma_{s_0,2}(x_1,x_2)=(F_{s_0,2}^D)^{-1}[F_{2} (x_2|x_1)]$. 

Given an original two-dimensional map $\vf=(f_1(x_1,x_2),f_2(x_1,x_2))$, the Doob effective map \eqref{fDDd} thus takes the form,
$\vf^D=(f_1^D(\tilde{x}_1,\tilde{x}_2),f_2^D(\tilde{x}_1,\tilde{x}_2))$ 
with
\begin{widetext}
\be
f_1^D(\tilde{x}_1,\tilde{x}_2)=\gamma_{s_0,1}\{f_1[\gamma_{s_0,1}^{-1}(\tilde{x}_1), \gamma_{s_0,2}^{-1}(\tilde{x}_1,\tilde{x}_2)]\}
\label{doob2_1}
\ee
\be
f_2^D(\tilde{x}_1,\tilde{x}_2)=\gamma_{s_0,2}\{f_1[\gamma_{s_0,1}^{-1}(\tilde{x}_1),\gamma_{s_0,2}^{-1}(\tilde{x}_1, \tilde{x}_2)],f_2[\gamma_{s_0,1}^{-1}(\tilde{x}_1), \gamma_{s_0,2}^{-1}(\tilde{x}_1,\tilde{x}_2)]\}
\label{doob2_2}
\ee
\end{widetext}
where, $\gamma^{-1}_{s_0,i}$ for $i=1,2$ are the two components
of the inverse function $\vg^{-1}_{s_0}$.

In the application of this methodology to specific examples in the next section, the density $\rho_{s_0}(\tilde{x}_1, \tilde{x}_2)$ will play a fundamental role, as it is needed for $\gamma_{s_0,1}(x_1)$ and $\gamma_{s_0,2}(x_1,x_2)$, themselves based on the cumulative distribution in Eqs.~(\ref{fs01D}) and (\ref{fs02D}). Such density $\rho_{s_0}(\tilde{x}_1, \tilde{x}_2)$ is computed as in Eq.~(\ref{sinv}) from the eigenfunctions solving Eq.~(\ref{eigprob}). Since the latter are estimated numerically, based on discretizing the phase space $I^d$, our results will be based on approximations that are to some extent uncontrolled (though in principle susceptible to arbitrary improvement\rev{; see our discussion of Ulam's method in Subsec.~\ref{subsec:tilted}}). On each occasion we have checked in multiple forms that the results are consistent with the expectations based on the exponential biasing of the observable of interest, see Eq.~(\ref{tiltdist}), once the discretization points are sufficiently close to one another. Yet the singular nature of some of the eigenfunctions in Eq.~(\ref{eigprob}), and the potential difficulties that may arise from it, have been pointed out recently, together with analytical approaches that avoid these pitfalls at least for some specific one-dimensional maps \cite{Monthus2024}.

\section{\label{sec:applications} Applications}
Having presented a theoretical framework for making rare events typical in $d$-dimesional chaotic maps, including explicit expressions for $d=2$, we here apply it to one observable of the two-dimensional tent map and two observables of Arnold's cat map. While the former is a non-invertible map, the latter is invertible, and both have been studied in the literature on chaotic dynamics. In each case, one needs to solve the eigenvalue problem (\ref{eigprob}), which is usually done numerically, in order to study the statistics for different tiltings contained in the SCGF $\theta(s)$, choose the tilting giving rise to the rare event of interest, $s=s_0$, and obtain the corresponding invariant density $\rho_{s_0}$ (\ref{sinv}). For this purpose, we employ a two-dimensional adaptation of the power method described in the Supplemental Material of Ref.~\cite{Gutierrez2023}, which is also similar to the one described in Ref.~\cite{coghi23a}, with the uniform distribution as initial condition, sometimes in combination with other methods to be discussed below. \rev{A detailed description of the steps that are followed, together with remarks about their computational cost, can be found in Appendix \ref{app:compu}.}

\subsection{Two-dimensional tent map}
\label{tentsec}

In a series of articles, the two-dimensional tent map was presented and studied at length~\cite{pumarino2013,pumarino2014, pumarino2015,Alves2017}. Its name stems from the many shared properties with its one-dimensional counterpart, which was analyzed using the $d=1$ version of the large-deviation methodology presented here (see Subsec.~\ref{1dim}) in Ref.~\cite{Gutierrez2023}. Some of these properties are: the existence of a unique invariant measure with respect to which the system is ergodic, a conjugacy relation to a one-sided shift in a subset of two symbols, and the existence of a dense set of periodic orbits. Together with global folding and stretching, they characterize a system that nicely lends itself to the application of the methodology previously developed, as some results can be obtained analytically with relative ease.

\begin{figure}[tp]
    \centering
    \begin{tikzpicture}
        \coordinate (A) at (0,0); 
        \coordinate (B) at (5,0); 
        \coordinate (C) at (2.5,3); 
        \coordinate (D) at (2.5,0); 
        \coordinate (T0) at (1.5, 0.75); 
        \coordinate (T1) at (3, 0.75); 
        \coordinate (E) at (0,3); 
        \coordinate (F) at (0.5,2); 

        \draw (A) -- (B) -- (C) -- cycle;
        \draw[thick, red] (C) -- (B) -- (D) -- cycle;

        \draw[dashed, red] (A) -- (E) -- (C);

        \node at (T0) {$\mathcal{T}_0$};
        \node at (T1) {$\mathcal{T}_1$};
        \draw [->, out=50, in=20] (T1) to (F) ;
    \end{tikzpicture}
    
    \begin{tikzpicture}
        \draw[step=1.0,red,,dashed] (0,0) grid (3,3);
        \coordinate (A0) at (0,0);
        \coordinate (A1) at (1,0);
        \coordinate (A2) at (1,1);
        \coordinate (A3) at (0,1);
        \draw (A0) -- (A1) -- (A2) -- (A3) -- cycle;

        \coordinate (B0) at (1,0);
        \coordinate (B1) at (2,0);
        \coordinate (B2) at (2,1);
        \coordinate (B3) at (1,1);
        \draw (B0) -- (B1) -- (B2) -- (B3) -- cycle;

        \coordinate (C0) at (2,0);
        \coordinate (C1) at (3,0);
        \coordinate (C2) at (3,1);
        \coordinate (C3) at (2,1);
        \draw (C0) -- (C1) -- (C2) -- (C3) -- cycle;

        \coordinate (C0) at (3,0);
        \coordinate (C1) at (4,0);
        \coordinate (C2) at (4,1);
        \coordinate (C3) at (3,1);
        \draw[thick, red] (C0) -- (C1) -- (C2) -- (C3) -- cycle;

        \coordinate (D0) at (4,0);
        \coordinate (D1) at (5,0);
        \coordinate (D2) at (5,1);
        \coordinate (D3) at (4,1);
        \draw[thick,red] (D0) -- (D1) -- (D2) -- (D3) -- cycle;

        \coordinate (E0) at (1,1);
        \coordinate (E1) at (2,1);
        \coordinate (E2) at (2,2);
        \coordinate (E3) at (1,2);
        \draw (E0) -- (E1) -- (E2) -- (E3) -- cycle;
        \coordinate (E0) at (2,1);
        \coordinate (E1) at (3,1);
        \coordinate (E2) at (3,2);
        \coordinate (E3) at (2,2);
        \draw (E0) -- (E1) -- (E2) -- (E3) -- cycle;
        \coordinate (E0) at (3,1);
        \coordinate (E1) at (4,1);
        \coordinate (E2) at (4,2);
        \coordinate (E3) at (3,2);
        \draw[thick, red] (E0) -- (E1) -- (E2) -- (E3) -- cycle;

        \coordinate (E0) at (2,2);
        \coordinate (E1) at (3,2);
        \coordinate (E2) at (3,3);
        \coordinate (E3) at (2,3);
        \draw (E0) -- (E1) -- (E2) -- (E3) -- cycle;

        \coordinate (T0) at (1.5,0.5); 
        \coordinate (T1) at (3.5, 0.5);
        \coordinate (F) at (0.5, 2.5);
        \draw [->, out=50, in=20] (T1) to (F) ;
        \node at (T0) {$\mathcal{T}_0$};
        \node at (T1) {$\mathcal{T}_1$};
        
    \end{tikzpicture}
    \vspace{0.2cm}
    \caption{Application of ${\bf K}$ \eqref{K} to the domain $\mathcal{T} = \tleft\cup \tright$ of the two-dimensional tent map $\vLambda$ (\ref{tent}). Both the original form of ${\bf K}$ (upper panel) and its discretization (lower panel) are displayed. To better illustrate the method, the size of each discretization cell is unrealistically large in the diagram. In our numerical results, the discretization is much finer, based on a grid of $\sim 5\cdot 10^3\times 5\cdot 10^3$ cells.}
    \label{fig:tent_diagram}
\end{figure}
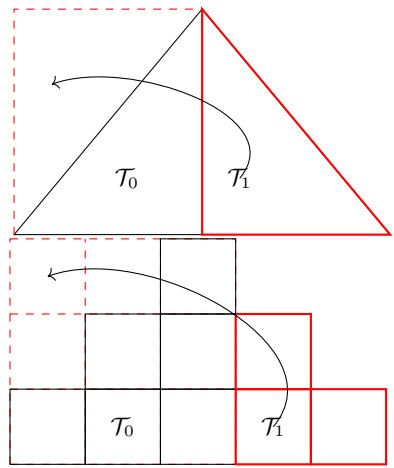

The two-dimensional tent map is defined as follows:
\begin{equation}
    \vLambda(x,y) = \begin{cases}
       (x+y, x-y) & (x,y)\in \tleft\\
       (2-x+y, 2-x-y) & (x,y)\in \tright, \\
    \end{cases}
    \label{tent}
\end{equation}
where the subsets of the plane $\tleft = \{(x,y)\in \mathbb{R}^2: 0\leq x \leq 1, 0 \leq y \leq x\}$ and $\tright = \{(x,y)\in \mathbb{R}^2: 1< x \leq 2, 0 \leq y \leq 2-x\}$ are illustrated in Fig.~\ref{fig:tent_diagram} (more about this figure later). In more compact form, we can define $\vLambda: \tcal \to \tcal$, where the phase space is $\tcal = \tleft\cup \tright$, as $\vLambda (x,y) = (1+y-|1-x|,1-y-|1-x|)$. The map has an invariant measure that is uniform in $\tcal$, and is expanding almost everywhere, with areas doubling at each iteration, as given by the Jacobian determinant $|J_{\vLambda}|=2$. The exception is the line segment at the boundary between $\tleft$ and $\tright$, $\mathcal{C}=\{(x,y)\in \mathbb{R}^2: x=1, 0\leq y \leq 1\}$, where $\vLambda$ is continuous but nondifferentiable.

Since every point in the triangle has exactly one preimage in $\tleft$ and another in $\tright$, the right eigenfunction
problem for an observable with local contribution $g_{\vLambda}$, based on Eqs.~(\ref{tiltop}) and (\ref{eigprob}),  reads
\begin{equation}
    \frac{1}{2}\left[e^{-sg_{\vLambda(\vz_1)}}r_s(\vz_1) + e^{-sg_{\vLambda(\vz_2)}}r_s(\vz_2)\right]= e^{\theta(s)} r_s(\vz),
    \label{eigdoubling}
\end{equation}
where for a given $\vz \in \tcal$, we have $\vz_1\in \tleft$ and $\vz_2 \in \tright$ such that $\vLambda(\vz_1) = \vLambda(\vz_2) = \vz$. The factor $1/2$ on the left-hand side is due to a division by $|J_{\vLambda}|$ arising from the Dirac delta in Eq.~(\ref{tiltop}).

Our discussion of rare events in the two-dimensional tent map \eqref{tent} will focus on an observable $A_N$ \eqref{obs} that quantifies the frequency of visits to $\tleft$ in a trajectory of $N$ steps. Its local contribution is the indicator function for $\tleft$, namely
\begin{equation}
    g_{\vLambda(x,y)} = \begin{cases}
        1, &(x,y) \in \tleft\\
        0, &(x,y) \in \tright. 
    \end{cases}
    \label{obstent}
\end{equation}
Since $e^{-sg(\vz_1)}=e^{-s}$ and $e^{-sg(\vz_2)} = 1$, an ansatz solution to \eqref{eigdoubling} would be
\begin{align}
    r_s(\vz) &= 1, \label{analySCGF1}\\
    \theta(s) &= \ln\left[ \frac{1}{2}(e^{-s} + 1)\right],\label{analySCGF2}
\end{align}
which holds for the one-dimensional tent map as well, with an equivalent observable. This is clearly a valid solution for the unbiased dynamics ($s=0$), as the SCGF satisfies $\theta(0) = 0$, yielding that the (uniform) invariant density is the right eigenfunction $r_s(\vz)$ with the largest eigenvalue of the Frobenius-Perron operator given by $e^{\theta(0)} = 1$. Postulating that the validity of Eqs.~(\ref{analySCGF1},\ref{analySCGF2}) extends to an interval of $s$ around the origin requires assuming that no level crossings in the spectrum of the tilted operator, which would signal the presence of a dynamical phase transition (DPT), are present for such $s\neq 0$. This assumption is vindicated by an excellent agreement with results obtained via two independent numerical methods, as we now explain.

\vspace{0.3cm}


As a first check on the validity of the ansatz on which our expression for the SCGF $\theta(s)$ \eqref{analySCGF2} depends, we numerically solve Eq.~(\ref{eigdoubling}) through the power method. The required discretization is streamlined by folding the map onto a unit square, see the upper panel of Fig.~\ref{fig:tent_diagram}. In practice, we employ the map $\tilde \vLambda  = \vK \circ \vLambda \circ \vK^{-1}$, given by the conjugation~\footnote{The motivation behind this conjugation is purely computational, since the matrix that contains the information about $r_s$ and $l_s$ is, by definition, rectangular. This prescription also significantly simplifies dealing with boundary issues.}
\begin{equation}
    {\bf K}(x,y) =
\begin{cases}
    (x, y), &(x,y) \in \tleft, \\
    (y, 2-x), &(x,y) \in \tright.
\end{cases}
\label{K}
\end{equation}
This assumes that $\tright$ does not include the line segment $y=2-x$, as otherwise those points and those in $y=x$, which belong to $\tleft$, would have the same images under ${\bf K}(x,y)$. In the discretized version of $\tcal$, as the phase space can be folded without overlaps, bijectivity is guaranteed; see the lower panel of Fig.~\ref{fig:tent_diagram}. 
Additionally, we obtain the SCGF $\theta(s)$ by means of the cloning algorithm, see Refs.~\cite{ giardina06a, tailleur07, laffargue13a}, based on tracking biased trajectories for long times in a scheme where the number of trajectories (``clones'') changes depending on the local contribution to the observable. 
Given the deterministic nature of the system, a small normally-distributed white noise $\varepsilon =10^{-8} \mathcal{N}(0,1)$ is added in each step to the $y$ coordinate to ensure good trajectory sampling, as in Refs.~\cite{tailleur07, Gutierrez2023}.

\begin{figure}
    \centering
    \includegraphics[width=\linewidth]{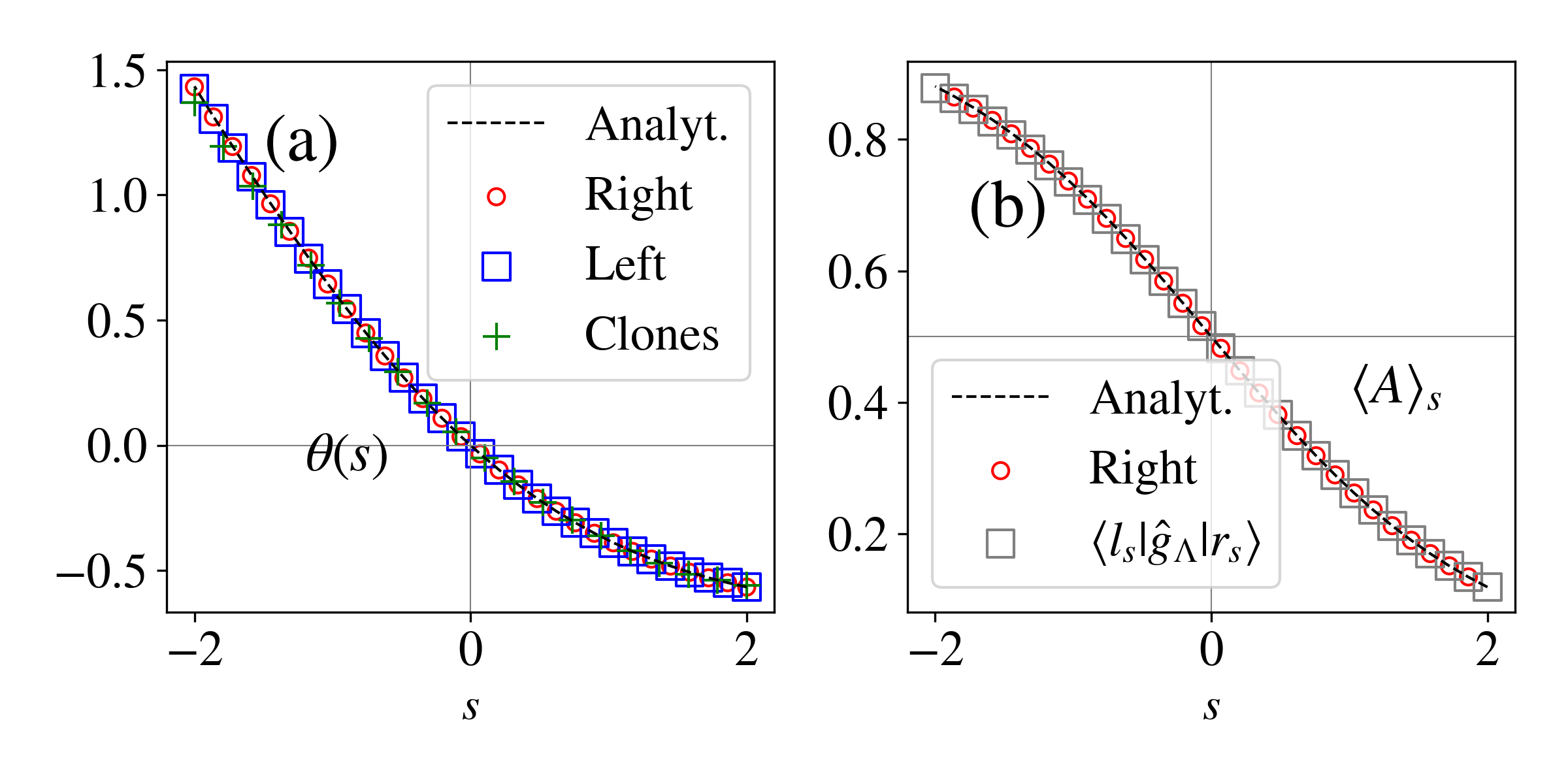}
    \vspace{-0.6cm}
    \caption{SCGF $\theta(s)$ and $s$-ensemble average $\la A \ra_s$ for the two-dimensional tent map, including exact results and approximate values based on different numerical methods. (a) SCGF analytical (ansatz) solution \eqref{analySCGF2}, SCGF solution of the right and left eigenproblems $(\ref{eigprob})$ based on the power method with $15$ iterations and a discretization of the unit square of $L\times L$ points with $L=2\cdot 10^3$, and cloning-algorithm estimates of SCGF based on $5\cdot 10^4$ clones and 100 steps are shown. (b) Tilted average $\la A \ra_s$ based on the derivative of the analytical solution, together with results based on the numerical derivative of the SCGF as obtained from the right eigenproblem solved by means of the power method, and the average 
    $\la A \ra_s = \bra{l_s}\hat{g}_{\vLambda}\ket{r_s}$, with both eigenfunctions also derived from the power method, are shown.}
    \label{fig:tent_SCGF}
\end{figure}

In Fig.~\ref{fig:tent_SCGF} (a), we compare the graph resulting from the analytical expression for the SCGF $\theta(s)$ in Eq.~(\ref{analySCGF2}), with the values yielded by the numerical methods just described,  namely, the power method, which is applied on the right and left eigenfunction problems (\ref{eigprob}), and the cloning algorithm. We stress that these methods solve approximately the eigenproblem (\ref{eigprob}) for any $s\neq 0$, without relying on arguments based on continuation from $s=0$ (and the absence of a DPT) as those implied in the exact ansatz solution in Eqs.~(\ref{analySCGF1}) and (\ref{analySCGF2}). The agreement is nearly perfect, with only slight deviations for the cloning-algorithm results at large $|s|$. 

In Fig.~\ref{fig:tent_SCGF} (b), we present the $s$-ensemble average $\la A \ra_s$ computed in three different ways: i) $\la A \ra_s = -\theta'(s)$, where the derivative is calculated on the analytical form of $\theta(s)$ given in Eq.~(\ref{analySCGF2}), ii) $\la A \ra_s$ obtained as (minus) the numerical derivative of the SCGF derived from the power method applied to the right eigenvalue-eigenfunction problem, iii) $\la A \ra_s$ as the average $\bra{l_s}\hat{g}_{\vLambda} \ket{r_s}$, based on Eqs.~\eqref{sinv} and \eqref{ave_s2}, with $l_s$ and $r_s$ derived from the power method as well. These results further confirm the validity of the ansatz used in the derivation of analytical results, and also highlight the accuracy of our numerical schemes in finding approximate solutions to the eigenproblem. 

Since the  SCGF in Eq.~\eqref{analySCGF2} is convex, $\theta''(s) =e^{-s}/(e^{-s}+1)^2> 0 $,  we can apply the Legendre transform [i.e.,~the inverse of the transformation \eqref{leg}] to obtain the LDF
\begin{equation}
    I(a) = a\ln\left( \frac{a}{1-a}\right) - \ln\left[ \frac{1}{2}\left(\frac{1}{1-a}\right)\right],
    \label{ratefundoubl}
\end{equation}
which will be of use below. Given that $a = -\theta'(s) = e^{-s}/(1+e^{-s})$, we conclude that the domain of $I(a)$ is $0 < a < 1$, as expected from an observable whose local contribution is an indicator function \eqref{obstent}. Note that the LDF can be rewritten as $I(a) = -S(a) + \ln 2$, where $S(a) = -a \ln a - (1-a) \ln (1-a)$ is the Shannon entropy of a Bernoulli trial with probability $a$ (i.e.,~the probability of visiting $\tleft$ in a long trajectory). Since the latter vanishes as $a\to 0^+$ or $a\to 1^-$, $\ln 2$ is the limiting value of $I(a)$ at both endpoints of its domain.

\begin{figure}
    \centering
    \includegraphics[width=\linewidth]{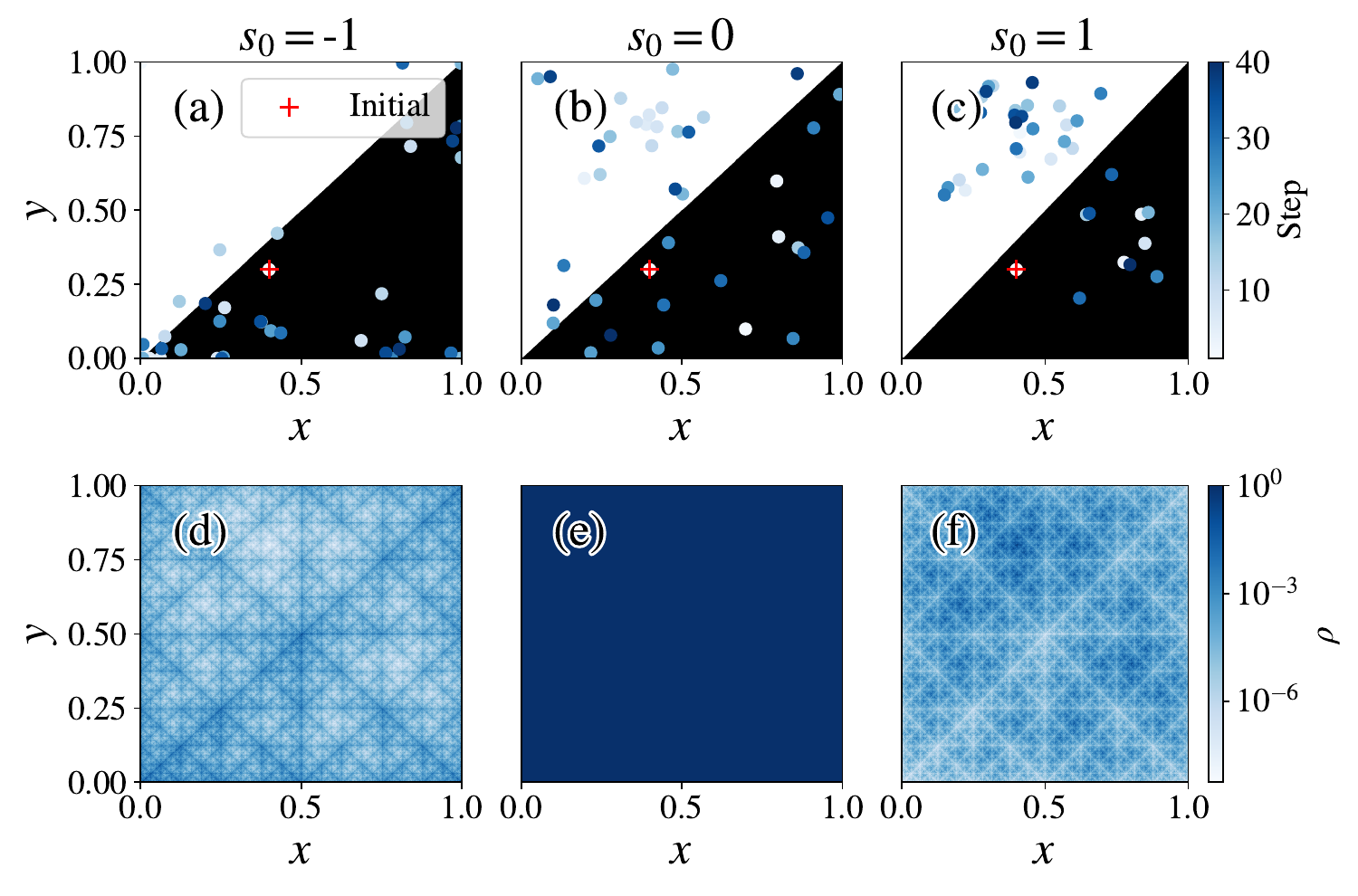}
    \caption{(a), (b), and (c): Representative trajectories of the two-dimensional tent map for the observable with local contribution (\ref{obstent}) corresponding to $s_0 = -1, 0$ (unbiased) and $1$, respectively. The filled circles are points along the trajectory colored by time step in the folded domain of the tent map (see Fig.~\ref{fig:tent_diagram}), with black (white) showing
    the $\tleft\, (\tright)$  region. The red cross indicates the initial position, (0.4, 0.3). (d), (e), and (f): Invariant densities $\rho_{s_0}$ \eqref{sinv} for $s_0= -1, 0$ and $1$, respectively, given in log scale and normalized by
    the maximum value in each case. }
    \label{fig:tent_3_plot}
\end{figure}

Having at our disposal the rare-event analysis displayed in Fig.~\ref{fig:tent_SCGF}, as well as the numerical solution (including the left eigenfunction $l_s$) of the eigenproblem (\ref{eigprob}), which is needed to build the tilted invariant density \eqref{sinv}, we can make rare events typical for the two-dimensional tent map. This is achieved by building the effective map (\ref{fDDd}), conjugate to $\vLambda$ through the transformation in Eqs.~\eqref{doob2_1} and \eqref{doob2_2}. The effective map is then used to generate typical trajectories corresponding to rare events in the original dynamics. In Fig.~\ref{fig:tent_3_plot}, we showcase representative trajectories --panels (a), (b) and (c)-- and the corresponding invariant densities --panels (d), (e) and (f)-- obtained for $s_0=-1$ [bias towards larger values of the observable; panels (a) and (d)], $s_0=0$ [natural dynamics; panels (b) and (e)] and $s_0=1$ [bias towards smaller values of the observable; panels (c) and (f)]. As $s$ is varied from zero, the trajectories move from a uniform distribution 
to spending an increasing amount of time in either of the regions considered in the definition of
$g_{\vLambda}$, $\tleft$ (for $s<0$) or $\tright$ (for $s>0$). In the cases displayed in the figure, this tilting translates into a shift of the average from $\la A \ra_{s_0=0} = 0.5$ to $\la A \ra_{s_0=-1} \approx0.73$ and $\la A \ra_{s_0=1}\approx0.27$; see also Fig.~\ref{fig:tent_SCGF} (b). In the last two cases, a fractal-like left eigenfunction $l_s$ (see the discussion in Ref.~\cite{Monthus2024}) results, in combination with the uniform right eigenfunction $r_s$ 
(\ref{analySCGF1}), in a fractal-like invariant density (\ref{sinv}) that underlies the biased statistics and is shown in panels (d), (e) and (f).

\begin{figure}
    \centering
    \includegraphics[width=\linewidth]{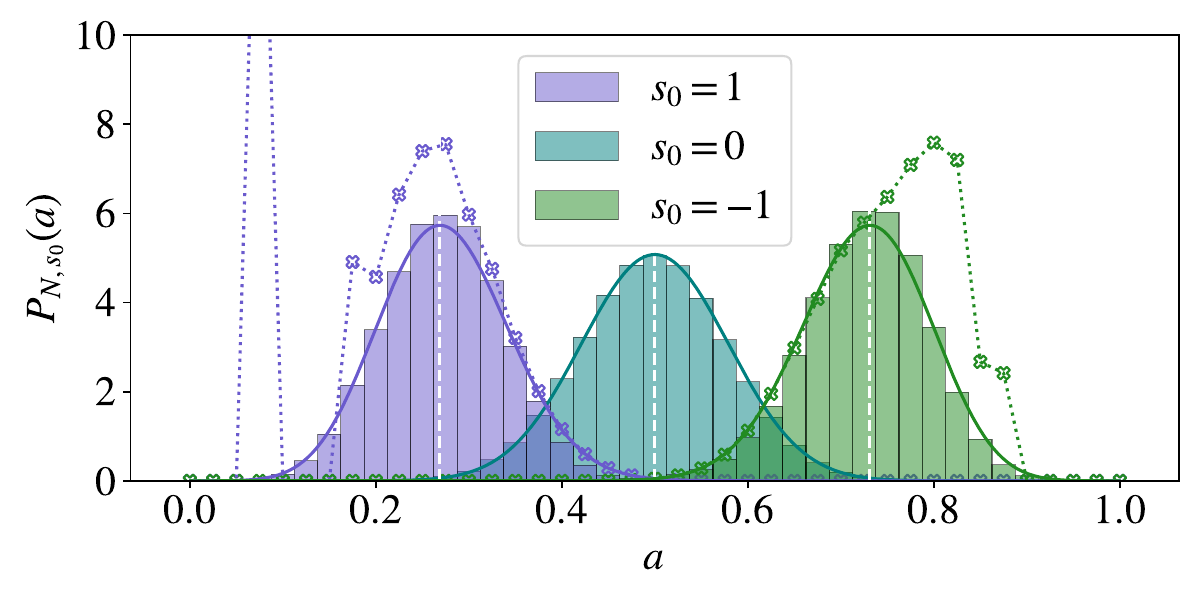}
    \caption{Histograms for the time-integrated observable with local contribution $g_{\vLambda}$ (\ref{obstent}) in the unbiased ($s_0=0$) and biased ($s_0=-1$ and $1$) two-dimensional tent map (\ref{tent}).  The solid lines represent the 
    analytical results arising from tilting the rate function (\ref{ratefundoubl}). The solid bars correspond to histograms obtained from the Doob effective map for each value of $s_0$. Dotted lines and crosses indicate the histogram obtained by a direct exponential reweighting of $P_{N}(a)$ yielding $P_{N,s_0} = e^{-s_0 N a} P_{N}(a)/Z_N(s_0)$, illustrating the poor sampling on the tails of the distribution. For the sake of visual clarity, the normalization $Z_N(s_0)$ employed for the rescaled histogram is calculated analytically, so that the scale is consistent across
    histograms. They are based on $5 \cdot 10^5$ trajectories of $N=40$ steps, both for the unbiased tent map and the Doob effective map (which coincide for $s_0=0$). Additionally, the white vertical dashed lines indicate the average $\bra{l_{s_0}} {\hat{g}}_{\vLambda}\ket{r_{s_0}}$ with eigenfunctions derived from the power method.}
    \label{fig:tent_Pa}
\end{figure}

The effective dynamics causes values of the time-integrated observable given by Eq.~(\ref{obstent}) that are exponentially suppressed in the natural dynamics to become typical statistics through an exponential bias that transforms the stationary distribution of the observable $P_N(a)$ into $P_{N,s_0}(a)$ (\ref{tiltdist}). If we were to directly access these events from the unbiased ($s_0=0$) distribution $P_N(a)$, those at the tails would be poorly
sampled, and obtaining $P_{N,s_0}(a)$ from an estimation of $P_N(a)$ and the direct exponential reweighting $P_{N,s_0}(a)= e^{-s_0Na} P_N(a)/Z_N(s_0)$ would require a sample that is exponentially large in $N$. To illustrate these difficulties and the efficiency of the large-deviation methodology in overcoming them, in Fig.~\ref{fig:tent_Pa} we show both the histogram derived from a sampling of the trajectories in the Doob effective
maps (solid bars), which naturally yield the distribution $P_{N,s_0}(a)$, and a sampling based on exponentially reweighting $P_N(a)$ in order to obtain $P_{N,s_0}(a)$ (crosses and dotted lines). Despite the large number of trajectories of $N=40$ steps included in the histograms ($5\cdot 10^5$), the discrepancy between the estimates is quite conspicuous, and it increases as one explores the exponentially suppressed values further into the tails of the distribution. Fig.~\ref{fig:tent_Pa} also displays a solid line, which shows the result of taking the analytical rate function $I(a)$ in Eq.~(\ref{ratefundoubl}), finding $I^D_{s_0}(a)=s_0 a+I(a)+\theta(s_0)$ for each $s=s_0$ under consideration (in this case, $s_0= -1$ and $1$), and then plotting the biased distribution $P_{N,s_0}(a)\sim e^{-N I^D_{s_0}(a)}$, as explained in Subsec.~\ref{doobgen}. This shows a very good agreement with the histograms that are empirically obtained from the trajectories of the Doob effective map, further highlighting the consistency of the approach.

\subsection{Arnold's cat map}

\begin{figure}
    \centering
    \includegraphics[width=\linewidth]{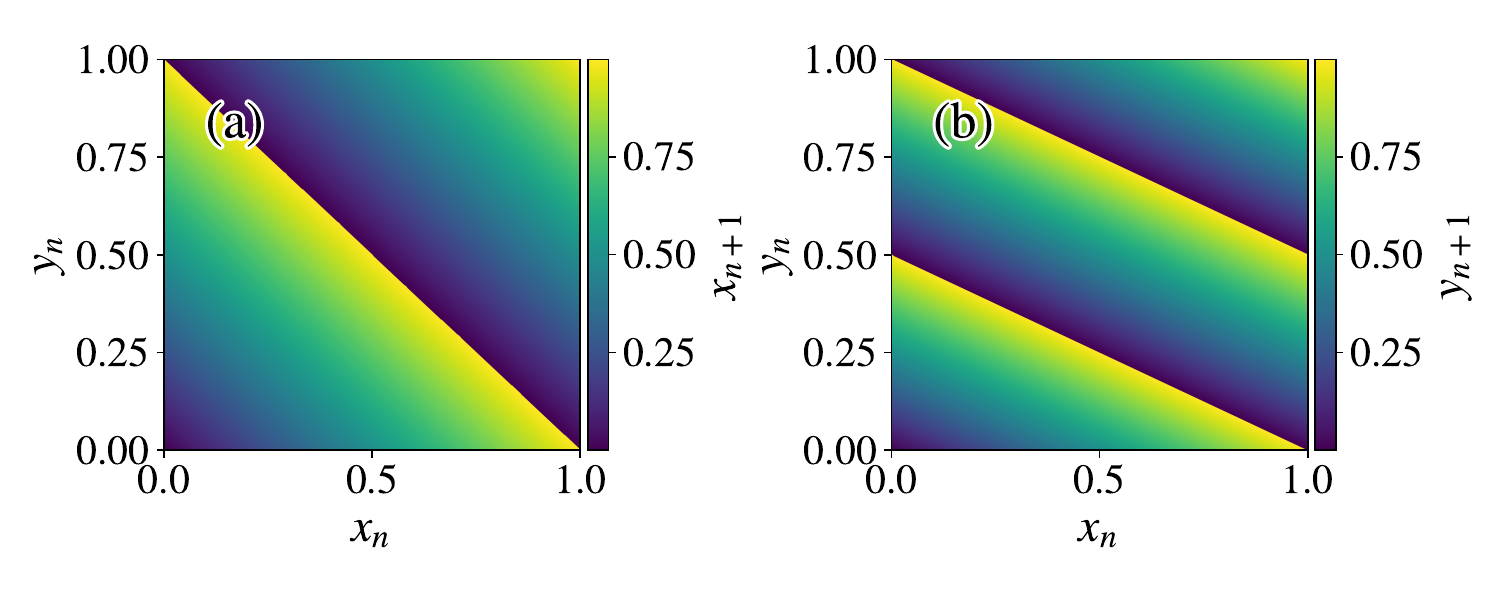}
    \caption{Colormaps corresponding to the graphs of the two components of Arnold cat's map \eqref{arnold}. (a) Map component corresponding to the $x$ variable, $x_{n+1} = (x_n+y_n) \mod 1$. (b) Map component corresponding to the $y$ variable, $y_{n+1} = (x_n+ 2 y_n) \mod 1$. }
    \label{fig:base}
\end{figure}

Arnold's cat map gives the discrete-time evolution of a state vector in $I^2=[0,1)\times[0,1)$ as $(x_{n+1},y_{n+1})={\bf c}(x_n,y_n)$, with
\begin{equation}
    {\bf c}(x_n,y_n)=[(x_n + y_n)\!\!\!\!\!\mod 1, (x_n + 2y_n)\!\!\!\!\!\mod 1].
    \label{arnold}
\end{equation}
It is a map from the torus $\mathbb{R}^2/\mathbb{Z}^2$ onto itself, its invariant measure is uniform, and it belongs to the K-automorphism family~\cite{Lasota1994}. The two components of the cat map are illustrated by color plots in Fig.~\ref{fig:base} to facilitate the comparison with the effective maps to be derived from it below. Two different observables of the map will be analyzed to gain insight into different aspects of the dynamics as reflected in its rare trajectories sustaining atypical statistics.  

\subsubsection{Observable A: Long time average of $(x+y)/2$}
\label{Acat}

We first focus on a time-integrated observable of the form \eqref{obs} with local contribution
\begin{equation}
    g_A(x,y) = (x+y)/2.
    \label{gA}
\end{equation}
The rare events of this observable, based on the tilted distribution \eqref{tiltdist}, have been previously studied in Ref.~\cite{Smith22}, which makes it a good candidate for a first application of the generalized-Doob-transform methodology to Arnold's cat map.
The most salient feature of the observable fluctuations is the presence of a first-order DPT located at $s\simeq \pm 4.7$, which will be addressed below.
\begin{figure}
    \centering
    \includegraphics[width=\linewidth]{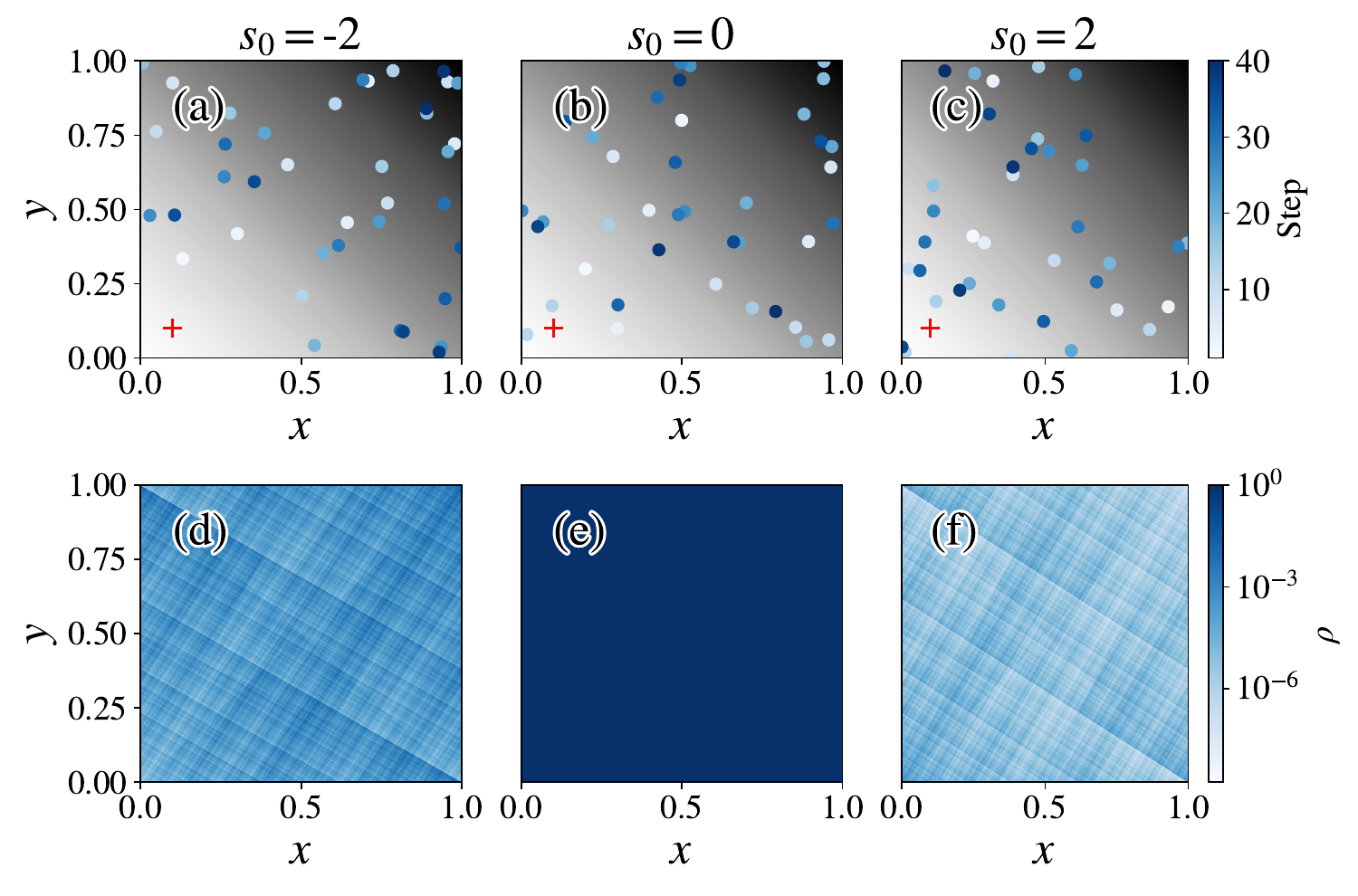}
    \caption{Representative trajectories [(a), (b) and (c)] and invariant densities [(d), (e) and (f)] as in Fig.~\ref{fig:tent_3_plot} but for Arnold's cat map with observable $A$, with tilting parameter $s_0 = -2$ [panels (a) and (d)], $s_0 = 0$ (unbiased) [panels (b) and (e)] and $s_0=2$ [panels (c) and (f)]. The local contribution $g_A$ \eqref{gA}
    is shown in greyscale from 0 (white) to 1 (black) in the upper panels. The initial position (red cross) is chosen to be (0.1, 0.1) in all three trajectories. }
    \label{fig:A_3_plot}
\end{figure}

\begin{figure}
    \centering    \includegraphics[width=\linewidth]{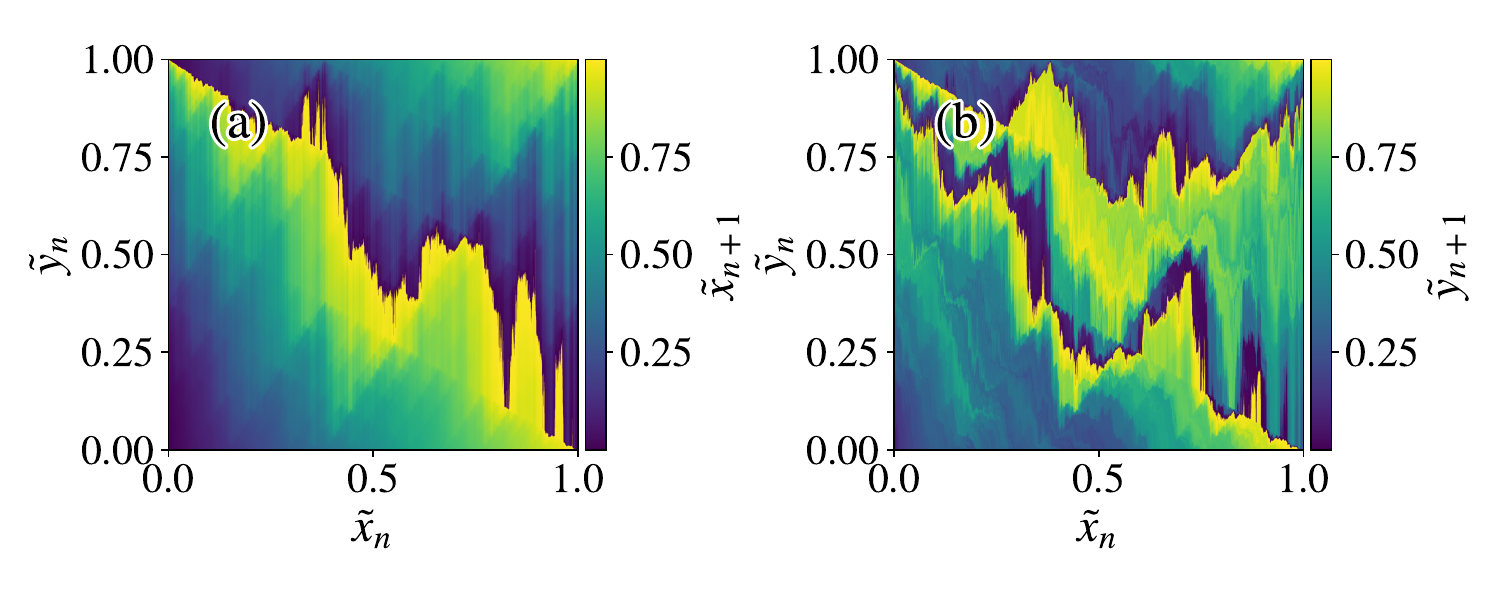} 
    \caption{Colormaps corresponding to the graphs of the two components of the  Doob effective map $\bm{c}_{s_0}^D(\tilde{x}_n,\tilde{y}_n)$ \eqref{fDDd} for Arnold's cat map with observable $A$ and tilting parameter $s_0=-2$. (a) Map component corresponding to the $\tilde{x}$ variable. (b) Map component corresponding to the $\tilde{y}$ variable. To be compared with the original map in Fig.~\ref{fig:base}.}
    \label{fig:A_doob}
\end{figure}

Before that, we consider representative trajectories obtained from the Doob effective map \eqref{fDDd} associated with the cat map, both with negative and positive tilting parameter values $s=s_0$, where $|s_0|< 4.7$ (away from the DPT), and including the unbiased dynamics $s_0=0$. Three such trajectories from the same initial position are shown in Fig.~\ref{fig:A_3_plot} (a), (b)  and (c). When the system 
is tilted towards a larger (smaller) value for the average $\left<A\right>_{s_0}$, see panel (a) [(c)] with $s_0<0$ ($s_0>0$), the trajectories concentrate in the top right (bottom left) corner of the domain, even for moderate $s_0$ very far from the DPT point. This is reflected in the average of the observable, which is displaced from $\left<A\right>_{s_0=0}=0.5$ to $\left<A\right>_{s_0=-2}\approx 0.58$ or $\left<A\right>_{s_0=2}\approx0.42$, respectively. For the sake of illustration, the Doob effective map for $s_0=-2$ is presented in Fig.~\ref{fig:A_doob}, which can be compared with the unbiased case in Fig.~\ref{fig:base}.  Just as in the one-dimensional scenario~\cite{Gutierrez2023}, the connection between the Doob map and the tilted statistics is far from obvious (which is expected for the long-time behavior of a chaotic map). Concerning the invariant
densities, which are shown for the same tilting parameter values $s_0$ in Fig.~\ref{fig:A_3_plot} (d), (e) and (f), since the map is invertible, 
both the right and left eigenfunctions are singular~\cite{Monthus2024}, making it numerically challenging to  
calculate $\rho_{s_0}$ \ref{sinv}. 

\begin{figure}[htpb]
    \centering
    \includegraphics[width=\linewidth]{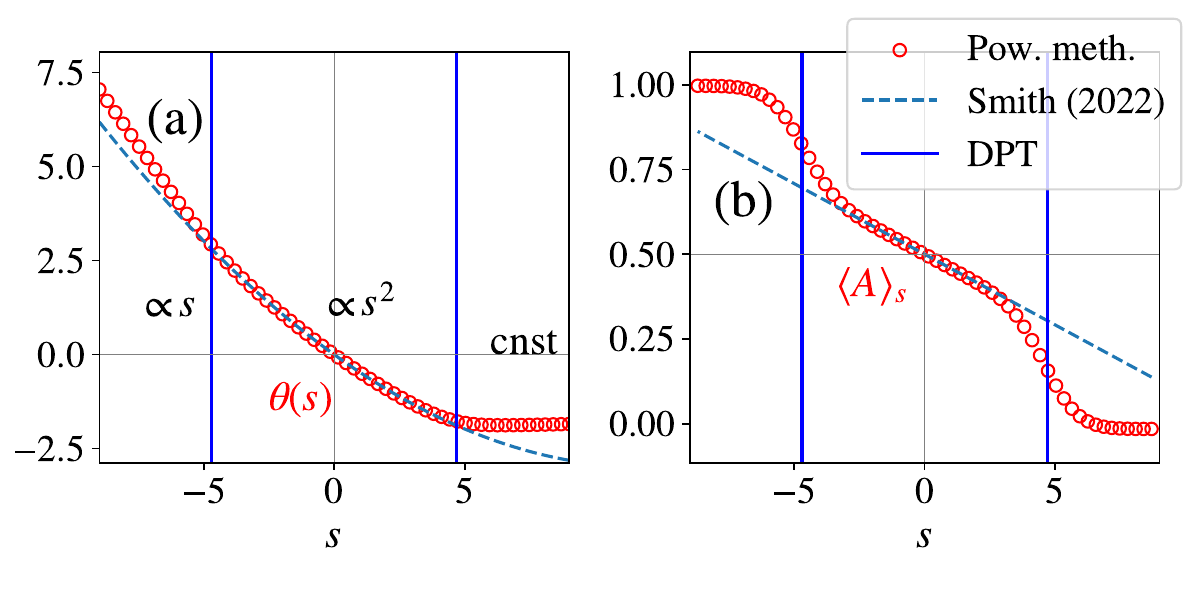}
    \caption{(a) SCGF $\theta(s)$ of Arnold's cat map for observable $A$, with local contribution $g_A$ (\ref{gA}), obtained from the right eigenfunction problem (\ref{eigprob}). Numerical results obtained from  $i=6$ iterations of the power method with a discretization of $L\times L$ points with $L=10^4$. (b) Observable average $\la A\ra_s=-\theta'(s)$ obtained from the numerical derivative as a simple finite difference. In both panels, analytical results from Ref.~\cite{Smith22}, known to be valid for $s\in [-4.7, 4.7]$, are displayed as blue dashed lines, and the position of the DPT $|s|\approx 4.7$ is highlighted by blue vertical lines.}
    \label{fig:A_SCGF_compar}
\end{figure}

We now move on to discuss the observable fluctuations as given by the SCGF $\theta(s)$, see Fig.~\ref{fig:A_SCGF_compar} (a), and the aforementioned DPT, which is linked to extreme trajectories concentrating around a fixed point. An approximation around $s=0$ yields (see Ref.~\cite{Smith22}):
\begin{equation}\label{eq:approx_A_theta}
    \theta(s) = -\frac{s}{2} + \frac{s^2}{48} + O(s^3).
\end{equation}
The validity of this expansion stops at a non-analyticity, namely the discontinuity in the first derivative of the SCGF, see Fig.~\ref{fig:A_SCGF_compar} (b), causing a jump in the $s$-ensemble average $\langle A \rangle_s = -\theta'(s)$ for $s\simeq \pm 4.7$. The abrupt change in the dynamics at the DPT point, as captured here by the power method appears smoother than in the calculations based on Monte Carlo sampling ~\cite{Smith22} due to the discretization cutoff at small scales required by our methodology; see Appendix \ref{app:DPT} for details. 

Since the SCGF $\theta(s)$ becomes linear for $|s| \gtrsim 4.7$, it is no longer strictly convex. Outside of the interval where the strict convexity of $\theta(s)$ holds, the one-to-one 
relation between $\theta(s)$ and $I(a)$ does not \cite{touchette09}, and some caution should be exercised in the interpretation of results, as the assumptions on which our methodology relies are not met. At the trajectory level, $-\theta'(s)=\la A\ra_s \approx 1$ for $s\lesssim-4.7$ translates into trajectories that spend a large fraction of time at phase-space points $(x,y)$ where both coordinates are extremely close to (but always less than) $1$, in the upper right corner of the domain $I^2$. Something similar happens for $s\gtrsim 4.7$, with $\la A\ra_s \approx 0$, now with trajectories concentrating at points $(x,y)$ where both coordinates lie very close (but are always greater than) $0$. In the torus geometry both sets of points lie very close to (yet on opposite sides of) the fixed point $(0,0)$, but the periodicity of boundaries is not reflected in the definition of the observable \eqref{gA}. The dynamics (as given by the Doob effective map) corresponding to such limiting cases, $|s| \gtrsim 4.7$,  yields trajectories (not shown) that concentrate around those extreme points most of the time, but in fact not permanently, as that would imply ergodicity breaking, which cannot arise from a topological conjugation. 
See Ref.~\cite{Gutierrez2023} for a more detailed discussion of the same point in the context of one-dimensional maps.

\subsubsection{Observable B: Checkerboard indicator}
\label{Bcat}

\begin{figure}
    \centering
    \includegraphics[width=\linewidth]{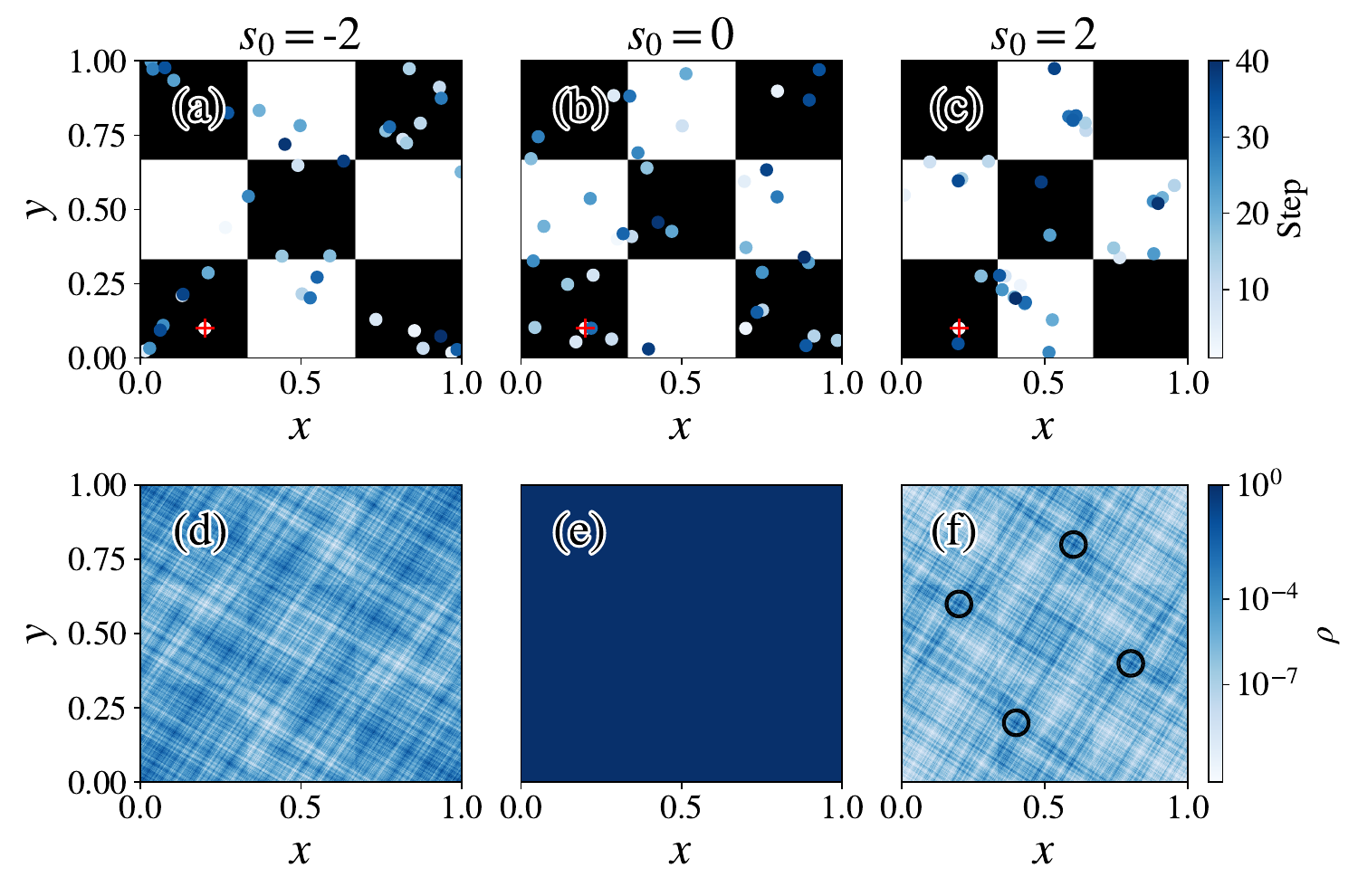}
    \caption{This figure contains analogous results, displayed with the same format, as Figs.~\ref{fig:tent_3_plot} and \ref{fig:A_3_plot} but for Arnold's cat map with observable $B$. 
    In (a), (b), and (c), black (white) shows $K$ ($I^2-K$), the region where the local contribution $g_B=1$ (0). The initial position (red cross) is chosen to be $(0.2, 0.1)$ to emphasize the bias in the dynamics.
    In (f), the period-2 points are represented as black circles.
    }
    \label{fig:B_3_plot}
\end{figure}

To conclude, we study the fluctuations of Arnold's cat map through another time-integrated 
observable, denoted as $B$, with probability distributed according to $P_N(b)$. Its local contribution is given by $g_B(x,y)=\mathbb{I}_K(x,y)$, where $\mathbb{I}_K$ assigns 
a value 1 to points $(x,y)$ in the region $K\subset I^2$ and 0 otherwise. This is an indicator function, similar to the one employed in the case of the tent map in Subsec.~\ref{tentsec}, but with support defined by a checkerboard-style configuration $K$. See an illustration in Fig.~\ref{fig:B_3_plot} [panels (a), (b) and (c)], where the 
black squares correspond to the subset of $I^2$ that makes up region $K$ (the results displayed in the figure will be discussed below).

\begin{figure}
    \centering
    \includegraphics[width=\linewidth]{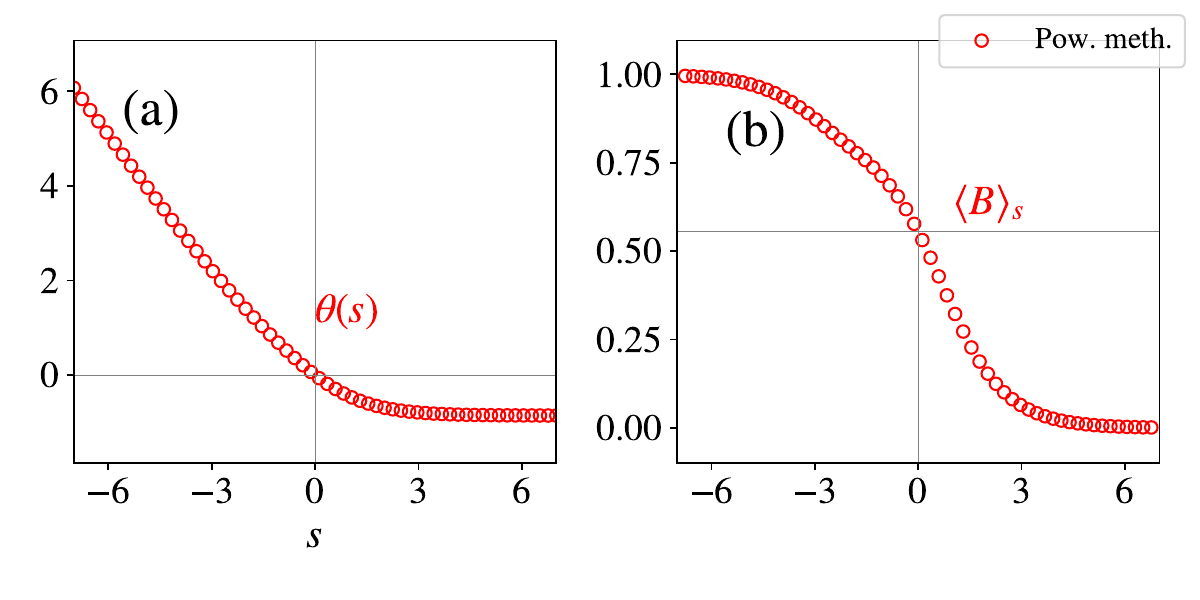}
    \caption{(a) SCGF $\theta(s)$ of Arnold's cat map for observable $B$, with local contribution $g_B$, obtained from the right eigenfunction problem (\ref{eigprob}) (b) Numerical derivative (based on a simple finite difference) of $\theta(s)$ for the right eigenfunction. Numerical results based on the power method with a discretization of $L\times L$ points with $L=4\cdot10^3$ and $i=6$ iterations.}
    \label{fig:B_SCGF_compar}
\end{figure}

The SCGF of Arnold's cat map conditioned on this observable is displayed in Fig.~\ref{fig:B_SCGF_compar} (a). It exhibits a rich behavior, which could be related to a dynamical crossover, as there is an abrupt change in slope in the derivative $\langle B\rangle_s = -\theta'(s)$ for negative $s$ as displayed in Fig.~\ref{fig:B_SCGF_compar} (b). 

As in the previous examples, the limiting values of $\la B \ra_s$ align with the maximum and minimum theoretical values for $B$, which are $0$ and $1$, see Fig.~\ref{fig:B_SCGF_compar} (b). The limit $\lim_{s\to-\infty} \la B \ra_s=1$ corresponds to trajectories 
confined inside of $K$. It turns out that these are trajectories that remain extremely
close to the fixed point most of the time (not shown). The opposite limit $\lim_{s\to \infty} \la B \ra_s=0$ arises because there are trajectories that spend an arbitrary amount of time away from $K$. In fact,  we find that the biasing favors 
trajectories that remain close to the period-2 orbits of Arnold's cat map, which arise from the initial points
\begin{equation}
    \left\{\left(\frac{1}{5}, \frac{3}{5}\right), \left(\frac{3}{5}, \frac{4}{5}\right), \left(\frac{2}{5}, \frac{1}{5}\right), \left(\frac{4}{5}, \frac{2}{5}\right)\right\},
\end{equation}
all of which are outside of $K$, as illustrated in Fig.~\ref{fig:B_3_plot} (f). In fact the limit $s\gg 0$ numerically results in an invariant density that concentrates precisely 
around those points.

\begin{figure}
    \centering
    \includegraphics[width=\linewidth]{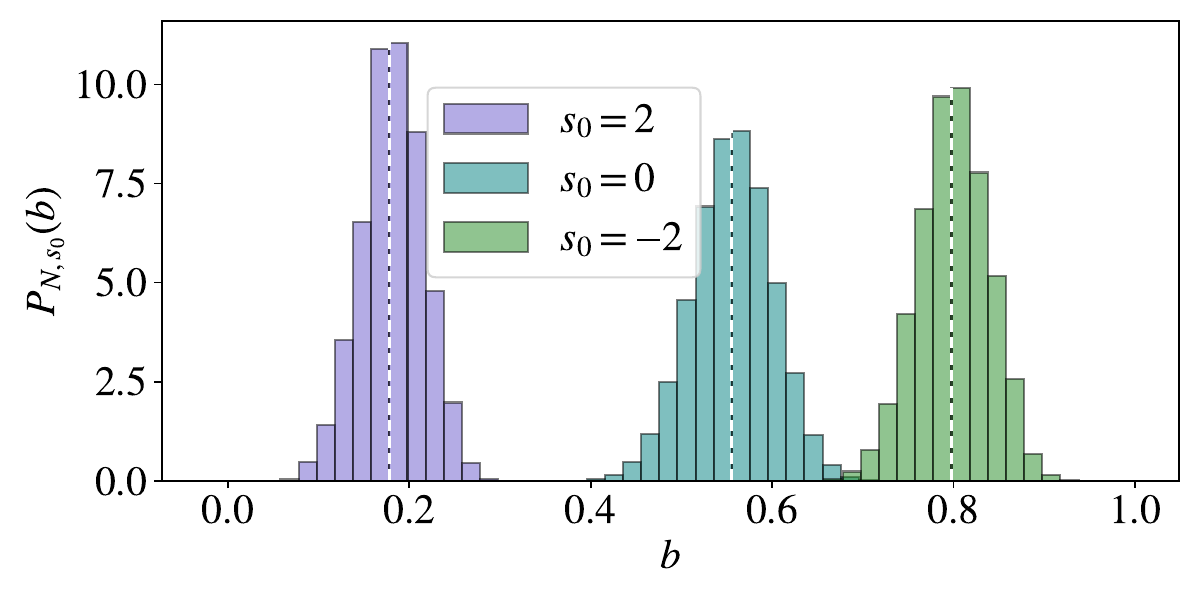}
    \caption{Histograms for the time-integrated observable with local contribution $g_B$ in the unbiased $(s_0=0)$ and biased ($s_0=\pm2)$ effective dynamics. Results based on $10^5$
    trajectories of $N=100$ steps by letting them evolve under 
    the Doob map $\bm{c}_s^D$ from their initial state. The dotted white vertical lines indicate the average, $\la B \ra_s=  \bra{l_s}\hat{g}_B\ket{r_s}$. }
    \label{fig:B_p_a}
\end{figure}

\begin{figure}
    \centering    \includegraphics[width=\linewidth]{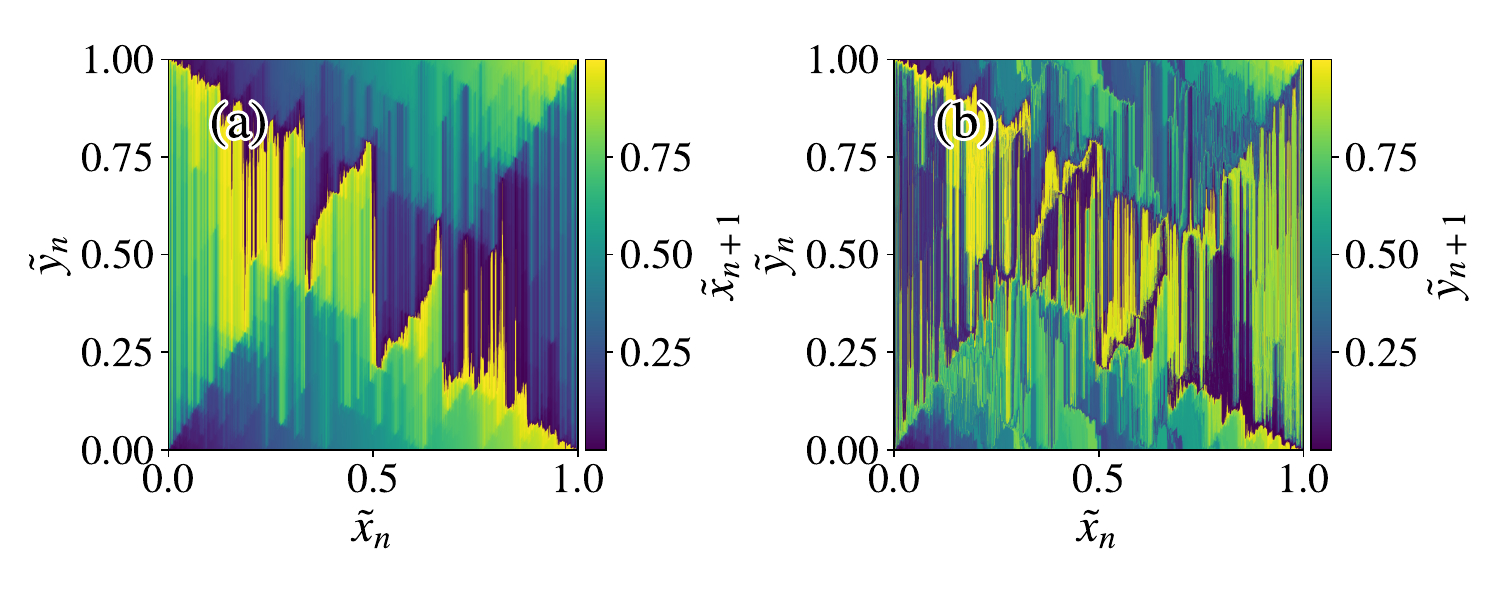} 
    \caption{Colormaps corresponding to the graphs of the two components of the  Doob effective map $\bm{c}_{s_0}^D(\tilde{x}_n,\tilde{y}_n)$ \eqref{fDDd} for Arnold's cat map with observable $B$ and tilting parameter $s_0=-2$. (a) Map component corresponding to the $\tilde{x}$ variable. (b) Map component corresponding to the $\tilde{y}$ variable. To be compared with the original map in Fig.~\ref{fig:base}.} 
    \label{fig:B_doob}
\end{figure}

Given the relatively steep slope displayed by $\theta(s)$, see Fig.~\ref{fig:B_SCGF_compar} (a), a moderate tilt results in extremely rare trajectories
with averages that are significantly displaced from $\la B\ra_{s_0=0} = 5/9\approx 0.56$, as shown in Fig.~\ref{fig:B_SCGF_compar} (b), to $\la B \ra_{s_0=-2}\approx 0.8$ or $\la B \ra_{s_0=2}\approx0.18$, for instance. Representative trajectories for such conditionings, $s_0 = -2, 0$ and $2$, are displayed in Fig.~\ref{fig:B_3_plot}, see panels (a), (b) and (c), respectively, the corresponding invariant densities being shown in the panels right below them in each case. The histograms for $b$ in the biased statistics and for the same values of $s_0$ are shown in Fig~\ref{fig:B_p_a}.  An illustration of the Doob map for $s_0=-2$ is given in  Fig.~\ref{fig:B_doob}, which is indicative of a very complex dynamical behavior.

\section{Conclusions}

We have presented a theoretical framework for making rare events typical in chaotic maps of any dimensionality, which extends the one-dimensional results of Ref.~\cite{Gutierrez2023} to a much wider variety of systems. The main ingredients are concepts and tools developed in the study of dynamical large deviations, including tilted distributions and the generalized Doob transform, in combination with the Rosenblatt transform based on conditional probabilities, which makes it feasible to consider maps of dimension $d>1$ in a framework that was previously only available in one dimension. Taken together, they allow for the construction of a map that is topologically conjugate to the original one, yet with the key difference that its stationary statistics (as given by the natural invariant density) sustains prescribed fluctuations that are extremely unlikely in the original dynamics.

Following a detailed explanation touching on conceptual, theoretical and numerical aspects associated with the methodology, we have next presented applications of our framework to two important two-dimensional maps, namely the two-dimensional tent map and Arnold's cat map. The results are relevant for control purposes, i.e.,\! out of a given map one finds another one with prescribed statistics of the observables of interest, but also provide a deeper understanding of the spectrum of fluctuations hidden in a given dynamics, including the existence and nature of  DPTs and the phases involved.

In the examples considered there is a recurring pattern, which was also observed in the one-dimensional setting \cite{Gutierrez2023}: the trajectories 
that dominate extreme behaviors for a given observable concentrate close to specific zero-measure sets, such as fixed points and periodic orbits, oftentimes inducing extreme changes in the statistics that may be
associated with DPTs.

The approximations introduced in our analysis, which are based on the use of discretizations of phase space and the convergence of numerical methods applied to them, are analogous to those present in the one-dimensional case  (see the Supplemental Material of \cite{Gutierrez2023}). While they are in principle susceptible to arbitrary improvement by an increase of the computational resources devoted to the power method, we consider it likely that the discussion in Ref.~\cite{Monthus2024} concerning the difficulties associated with the singularity of certain eigenfunctions of one-dimensional maps is relevant in higher dimensions as well. In this regard, the combination of forward deterministic dynamics with the backward stochastic dynamics in the analyses of irreversible maps proposed in that reference may be also important whenever exact (or high-precision) results are the goal in this new context.

Several interesting developments are expected to arise from this work. For instance, the adaptation of the finite-time generalized Doob transform \cite{chetrite15a,garrahan16} in the context of chaotic maps, which would allow us to obtain topologically-conjugate maps that typically yield prescribed observable statistics within finite time intervals. And, of course,  the extension of our framework to continuous-time flows given by (systems of) ordinary differential equations remains an alluring possibility.

\begin{acknowledgments}

The authors thank Naftali R.~Smith for insightful remarks on an earlier preprint version. The research leading to these results has received funding from the I+D+i grants PID2023-149365NB- I00, PID2020-113681GB-I00, PID2021-128970OA-I00, PID2021-123969NB-I00, PID2024-159024NB-C21, and C-EXP-251-UGR23, funded by MICIU/AEI/10.13039/501100011033/, ERDF/EU, Junta de Andalucía - Consejería de Economía y Conocimiento. YGK acknowledges financial support from PREP2023-001684 funded by MCIU/AEI/10.13039/501100011033 and the FSE+. We are also grateful for the computing resources and related technical support provided by PROTEUS, the supercomputing center of Institute Carlos I in Granada, Spain.
\end{acknowledgments}

\appendix

\section{Convergence of the SCGF near DPTs}
\label{app:DPT}

Whenever trajectories associated with a fluctuation are localized in phase space, the iterative procedure behind the power method that we employ (a two-dimensional adaptation of the algorithm described in the Supplemental Material of Ref.~\cite{Gutierrez2023}) may fail to provide accurate results. 
\begin{figure}[htpb]
    \centering
    \includegraphics[width=0.95\linewidth]{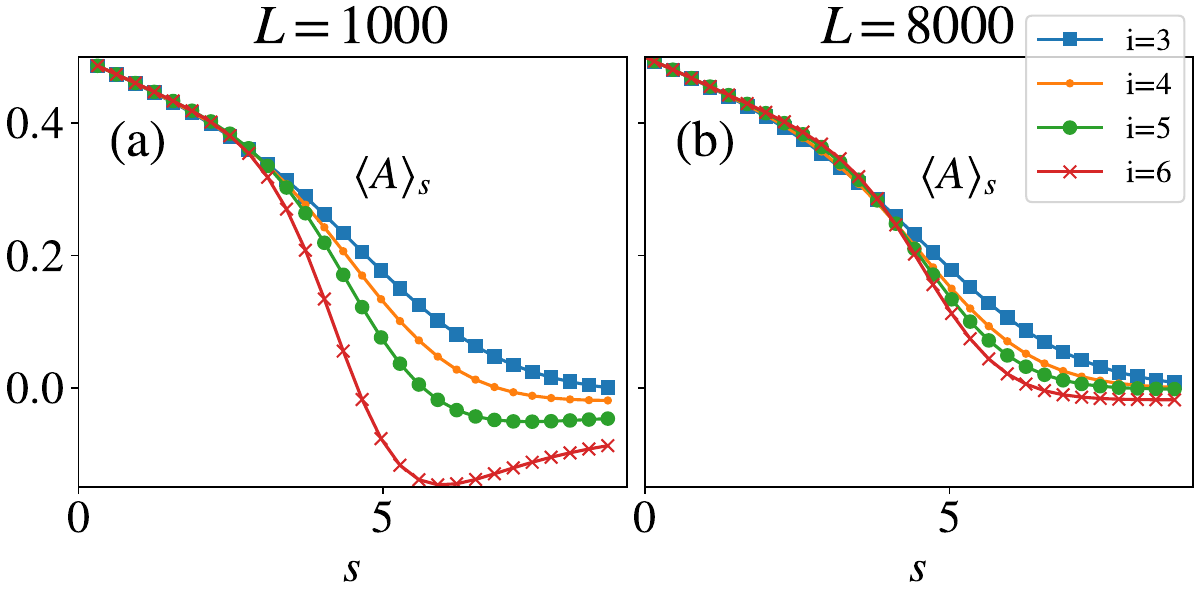}
    \caption{Tilted average $\la A \ra_s$ for the observable $A$ of Arnold's cat map for different power-method iterations $i = 3,4,5$ and $6$ obtained from the numerical derivative of the right eigenfunction problem. Results based on a discretization of $L\times L$ sites, with  $L=10^3$ (a) and $L=8\cdot10^3$ (b). The existence of a DPT is discussed in Subsec.~\ref{Acat}.}
    \label{fig:SCGF_conv}
\end{figure}

One such instance is
presented in Fig.~\ref{fig:A_SCGF_compar}, where the first-order DPT associated with observable $A$ of Arnold's cat map cannot be
faithfully characterized because of the inherent concentration of trajectories in small phase-space regions. To further illustrate the situation, Fig.~\ref{fig:SCGF_conv} (b) shows results for $\la A \ra_s$ corresponding to different iterations with a discretization grid of $L\times L$ sites with $L=8\cdot10^3$. Note that the results are displayed for $s>0$, and indeed as $s$ increases, the trajectory gets more localized in the bottom left corner of phase space, closer to $(0,0)$, eventually leading to the DPT. While there is no question that the visualization of the non-analyticity underlying the DPT improves as the number of iterations increases, the results are not as compelling as would be desirable.

This issue appears to be strongly related to the phase-space discretization. In practice, the maximum number of iterations that can be applied with the power method before encountering numerical problems (here $i=6$) is directly related to the size of the discretization cells. To illustrate this, in Fig.~\ref{fig:SCGF_conv} (a) we show analogous results for $\la A \ra_s$ but with just $L=1000$. The convergence of the method is compromised when a certain number of iterations is reached [$i=6$, for which $L=8000$ in panel (b) does not seem to be affected by such problems], resulting in large negative (hence non-physical) values for the average. 

Furthermore, one of the obvious effects of phase-space discretization is the impossibility of representing truly aperiodic orbits, as trajectories in a deterministic system with a finite number of states are necessarily periodic. How well our numerical schemes approximate aperiodic orbits for sufficiently long times crucially depends on the size of the discretization cell. In the case of trajectories localized within very small regions of phase space, such as those underlying the DPT under discussion, discretization cells of extremely small size would be needed to resolve the evolution. 

\section{\rev{Computational implementation}}
\label{app:compu}

\rev{Next, we lay out the algorithm that implements our methodological framework to make rare events typical, which we have followed to obtain the results discussed in Sec.~\ref{sec:applications}. For illustration purposes, we limit ourselves to the 2D case, as the maps considered in that section are bidimensional. This approach is also consistent with the freely available implementation at \url{https://github.com/yllari/LDT_2D_maps}. Generalizing it to higher dimensions should not pose any serious difficulty at a conceptual level (for aspects related to the computational cost, see below).} 

\begin{figure}[htpb]
    \centering
    \includegraphics[width=\linewidth]{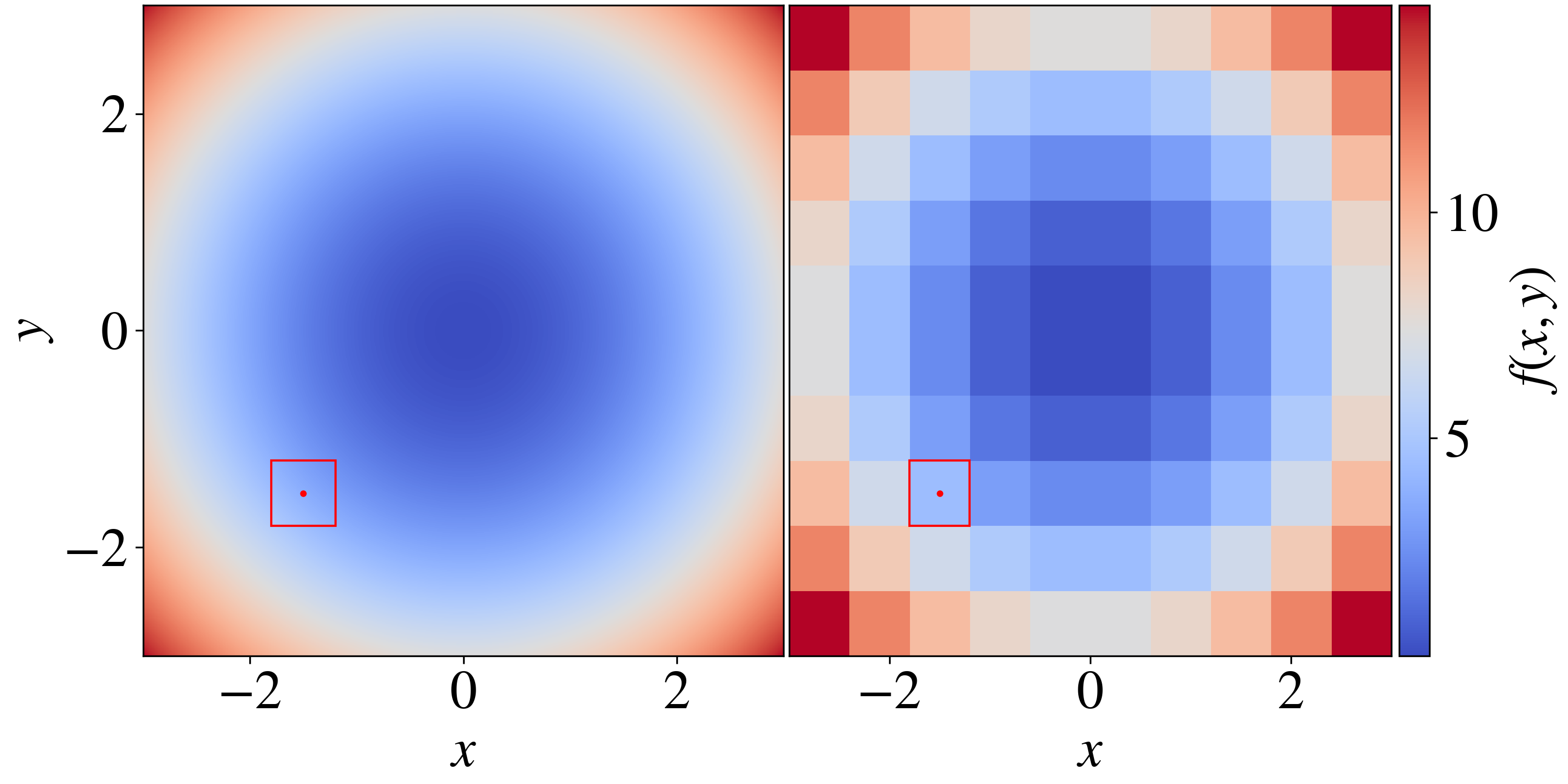}
    \caption{Example of discretization scheme used for the algorithm implementation applied to the function $f(x,y)=x^2 + y^2$. The value assigned to each cell is the central one.}
    \label{fig:2d_discret}
\end{figure}

\rev{In this implementation, the phase-space (i.e.\! a rectangular region in the plane) is discretized, and densities and other functions defined over it are stored in $N\times M$ matrices. The first index (row number, $i=0,1,\ldots, N-1$) encodes the discretization binning along the $y$ axis (from top to bottom), and the second (column number, $j=0,1,\ldots, M-1$) that along the $x$ axis (from left to right), both with equally-spaced bins. This is illustrated in Fig.~\ref{fig:2d_discret}, where the discretization scheme is applied to a given function $f(x,y)$. The highlighted square on the left panel corresponds to the discretization cell $[7,2]$ in the discretization on the right panel. (The area of the cell has been taken to be unrealistically large for visibility purposes.) While $f(x,y)$ obviously varies within the square, the assigned value of the discretization $f_{7,2}$ is taken to be that of $f(x,y)$ at the centre of the cell. Integrals over two-dimensional space are thereby approximated as sums,}
\begin{equation}
    \int f(x,y)dx dy \rightarrow  \sum_{i=0}^{N-1} \sum_{j=0}^{M-1} f_{i,j} \Delta x \Delta y,
\end{equation}
\noindent \rev{where $\Delta x$ ($\Delta y$) is the size of the bins along the $x$ ($y$) axis. }

\rev{In fact, while the discretization as described here is valid for scalar-valued functions, such as densities $\rho(x,y)$, the chaotic maps that we study are vector valued functions with two components, $\vf(x,y) = (f_1(x,y), f_2(x,y))$. Such discretization must then be applied separately on each of the two components $f_1(x,y)$ and $f_2(x,y)$, resulting in a three-dimensional array of dimensions $2\times N \times M$ (with the first index choosing the component). This is done as follows:}
\\

\begin{algorithmic}[1]
    \Procedure{\texttt{eval2D}}{array f, array X}
    \textcolor{commentgray}{\Comment $X$ is a $2\times N \times M$ array }
    \State $G \gets$ A grid that defines the binning on the coordinate space
    \State $N, M \gets \text{size}(X[0])$
    \State $I \gets \text{index2D}(X, G)$ 
    \For {$i \in [0, N[$}
        \For {$j \in [0, M[$}
            \State $R[0,i,j] \gets f[0,I[1,i,j], I[0,i,j]]$
            \State $R[1,i,j] \gets f[1,I[1,i,j], I[0,i,j]]$
        \EndFor
    \EndFor
    \State \Return R
    \EndProcedure
\end{algorithmic}
\rev{The evaluation of scalar functions [such as $\rho(x,y)$], follows from the simplification of the previous algorithm:}

\begin{algorithmic}[1]
    \Procedure{\texttt{eval}}{array f, array X}
    \State $G \gets$ A grid that defines the binning on the coordinate space
    \State $N, M \gets \text{size}(X[0])$
    \State $I \gets \text{index2D}(X, G)$ 
    \For {$i \in [0, N[$}
        \For {$j \in [0, M[$}
            \State $R[i,j] \gets f[I[1,i,j], I[0,i,j]]$
        \EndFor
    \EndFor
    \State \Return R
    \EndProcedure
\end{algorithmic}
\rev{
These two functions are all that is needed to implement the Frobenius-Perron operator and its adjoint (either tilted or not). As an illustrative example, we showcase the resulting discretized tilted Frobenius-Perron operator~\eqref{tiltop}; other implementations follow analogous steps.
}
\begin{algorithmic}[1]
    \Procedure{\texttt{tiltedFP}}{array $\rho$, array $g$, array $f^{-1}$, array $|J_f|$,
    array $X$, float $s$}
    \State $G \gets$ A grid that defines the binning on the coordinate space
    \State $N, M \gets \text{size}(X[0])$
    \State $Z \gets $ \texttt{eval2D}$(f^{-1}, X)$ \textcolor{commentgray}{\Comment{If more than one preimage exists, this would be repeated for each of them}}
    \State $|J_f|_\text{inv} \gets \text{\texttt{eval}}(|J_f|, Z)$
    \State $\rho_\text{inv} \gets \text{\texttt{eval}}(\rho, Z)$
    \State $g_\text{inv} \gets \text{\texttt{eval}}(g, Z)$
    \For {$i \in [0, N[$}
        \For {$j \in [0, M[$}
            \State $\rho_{{L}_s}[i,j] \gets e^{-s\cdot g_\text{inv}[i,j]}\rho_\text{inv}[i,j]/|J_f|_\text{inv}[i,j]$ \textcolor{commentgray}{\Comment{Same here, each preimage would be added}}
        \EndFor
    \EndFor
    \State \Return $\rho_{L_s}$
    \EndProcedure
\end{algorithmic}

\rev{We note that the algorithmic complexity of applying \texttt{tiltedFP} is $O(2N\times M)$, corresponding to the evaluation of the two components of the inverse function $f^{-1}$. To explore the scalability of this approach to $d$ dimensions later on, we explicitly include the dimensionality factor $2$. When this routine is called for the invariant density calculation in the power method, the scalability is $O(2i N\times M)$, where $i$ is the number of iterations.}

\rev{Once the invariant density of interest, Eq.~(\ref{sinv}), has been constructed, we apply the Rosenblatt transform to it. Marginalization and cumulative integrals, which yield CDFs, become sums and cumulative sums along a specific axis, respectively. The cumulative probability distributions of interest are calculated using the routine \texttt{Cumul},} 
\\
\begin{algorithmic}[1]
    \Procedure{\texttt{Cumul}}{array $\rho$, float $\varDelta x$, float $\varDelta y$}
        \State $N, M\gets \text{size($\rho$)}$
        \State $\rho_1 \gets$ Sum($\rho$, axis=0)$\times \varDelta y$

        \State $F_{1} \gets \text{CumulativeSum}(\rho_1)\times \varDelta x$

        \For{$i \in [0,N]$}
            \For {$j\in [0,M]$}
                \State $\rho_2[i,j] \gets \rho[i,j]/ \rho_1[j]$ \textcolor{commentgray}{\Comment same $x$ coordinate division}
            \EndFor
        \EndFor \textcolor{commentgray}{\Comment Example of consistent treatment of coords. }

        \State $F_{2} \gets \text{CumulativeSum}(\rho_2, \text{axis}=0)\times \varDelta y$
        \State \Return $F_1, F_{2}$
    \EndProcedure
\end{algorithmic}
\vspace{0.2cm}
\rev{
As in the previous piece of code, the complexity grows as $O(2N\times M)$, where the factor 2 is again an explicit dimensional factor. While $N\times M$ arises from the number of operations inside the for loop and inside the CumulativeSum functions, the dimensionality factor is determined by each of the cumulative sums performed.}

\rev{
The previous routine allows us to obtain the cumulative distribution functions [Eqs.~\eqref{f01D},~\eqref{f02D}, and~\eqref{fs01D},~\eqref{fs02D}],  which we next use to calculate the function $\vg$ giving the topological conjugacy, see Eqs.~(\ref{doob2_1}) and (\ref{doob2_2}).} 
\\
\begin{algorithmic}[1]
    \Procedure{\texttt{Gamma}}{array $\rho$, array $\rho_s$, float $\varDelta x$, float $\varDelta y$}
        \State $F_{1}, F_{2} \gets$ \texttt{Cumul}$(\rho, \varDelta x, \varDelta y)$
        \State $F^D_{s_0,1}, F^D_{s_0,2} \gets$ \texttt{Cumul}$(\rho_s, \varDelta x, \varDelta y)$

        \State $N, M\gets \text{size($\rho$)}$

        \State $F_{1,\text{coord}} \gets$ Mesh($F_1$) \textcolor{commentgray}{\Comment{Create grid for coordinates, one $F_1$ row for each $y$}}
        \State $F_{1,\text{coord}}^D \gets$ Mesh($F_1$) 
        
        \State $\tilde{I}[0]\gets \text{CondIndex2D}(F_{1,\text{coord}} , F^D_{1.\text{coord}})$  \textcolor{commentgray}{\Comment{Discrete inverse. Index values in bins respecting coords.}}

        \State $\tilde{X}_0\gets \tilde{I}[0]$

        \State $\tilde{I}[1]\gets \text{CondIndex2D}(F_{2}, \text{eval}(F^D_{s_0, 2}, [\tilde{X}_0, \mathbb{I}])$ \textcolor{commentgray}{\Comment{$\mathbb{I}$ here means no eval. in $y$}}

        \State $\tilde{X} \gets \text{IndexToCoord}(\tilde{I})$ \textcolor{commentgray}{\Comment $\gamma, \gamma^{-1}$ function}

        \State \Return $\tilde{X}$ 
    \EndProcedure
\end{algorithmic} 
\vspace{0.2cm}
\rev{In \texttt{Gamma}, the most important step corresponds to the correct inverse calculation, which amounts to the indexing of values in a certain set of bins following the prescription given in Sec.~\ref{doob_sec} [for $F_2^{-1}(x_2|x_1)$~\eqref{fs02D}, the inverse is calculated column-wise, which corresponds to fixed $x$]. The complexity for this two-dimensional implementation is $O(2 N^{} \log M)$, where the $2$ factor stems from the dimensionality. The $N\log M$ complexity factor arises from the discrete inverse search, which is performed as a binary search over a sorted array of length $M$ in each of the $N$ rows. Again, this is performed twice, once per dimension, because the system is two-dimensional, hence the factor $2$.}

\rev{Finally, we obtain the Doob map by applying the topological conjugation in Eq.~(\ref{fDDd}) as follows:}
\\
\begin{algorithmic}[1]
    \Procedure{Calculate conjugacy}{array $X$, array $\rho$, array $\rho_s$, float $\varDelta x$, float $\varDelta y $}
        \State $\gamma \gets $\texttt{Gamma}($\rho$, $\rho_s$, $\varDelta x$, $\varDelta y$) \textcolor{commentgray}{\Comment $\rho, \rho_s$ arrays evaluated on the grid $N\times M$}
        \State $\gamma^{-1}\gets $\texttt{Gamma}($\rho_s$, $\rho$, $\varDelta x$, $\varDelta y$)
        \State $f^D_s \gets \text{eval2D}(\gamma, \text{eval2D}(f,\text{eval2D}(\gamma^{-1},X )) )$
    \EndProcedure
\end{algorithmic}
\vspace{0.2cm}

\rev{
To complete our algorithmic complexity analysis, we expand our discussion to $d$-dimensional systems. For simplicity, we limit ourselves to the case in which the number of intervals per dimension is equal, i.e., the arrays will be of size $N^d$. The scalability of the methodology encompasses both memory requirements and algorithmic complexity. On one hand, memory storage grows as $dN^d$ floating-point numbers (including every component of the map ${\bf f}$) plus a fixed number of $N^d$ matrices to encode each eigenfunction (left, right, and invariant density) and other intermediate calculations. On the other hand, the number of operations, i.e., the complexity of the algorithm, is O($dN^d$). To elaborate on this, we examine each step of the procedure. First, we note that the $d$-dimensional algorithmic complexities of the different procedures are:
}
\begin{itemize}
    \item \texttt{EvalDD} (for $d$ dimensions): $O(d N^d)$ 
    \item \texttt{Cumul}: $O(d N^d)$
    \item \texttt{Gamma} (calls \texttt{Cumul}): $O(d N^d + dN^{d-1}\log N) \sim O(dN^d)$
\end{itemize}
\rev{\noindent where the complexity for $d$-dimensional systems is calculated analogously to the 2D implementations (one can extrapolate the calculations from previous paragraphs). On this list, the \texttt{Gamma} procedure complexity is driven by the call to \texttt{Cumul}, indicated by the combination of logarithmic and exponential scalings. From these results, it follows that the complexity of the conjugacy is that of the call to \texttt{Gamma} and, in turn, \texttt{Cumul}: $O (dN^d)$ (see the pseudocode corresponding to the conjugacy procedure). Similarly, the invariant density calculation amounts to $O(i d N^d)$ given the $i$ iterations of the power Method on \texttt{EvalDD}. In conclusion, the algorithmic complexity corresponds to $O(d N^d)$ and is determined equally by the Frobenius-Perron iteration in the power method and the marginal and conditional probabilities calculation used in the conjugacy.}


\bibliography{referenciasDoobChaotic}

\end{document}